\documentclass{aa}  

\usepackage{graphicx}
\usepackage{txfonts}
\usepackage{color}
\usepackage{bm}
\usepackage{amsmath}

\usepackage{hyperref}
\hypersetup{
    colorlinks,
    linkcolor={blue},
    citecolor={blue},
    urlcolor={blue}
}
\newcommand{\pd}[2]{\frac{\partial #1}{\partial #2}}
\newcommand{\dd}[2]{\frac{\mathrm{d} #1}{\mathrm{d} #2}}
\newcommand{\diff}{\mathrm{d}}
\newcommand{\im}{\mathrm{i}}
\newcommand{\cs}{c_\mathrm{s}}

\newcommand{\newtext}[1]{\ifmmode\bm{ #1}\else\textbf{#1}\fi}

\begin{document}

   \title{Probing the eccentricity in protostellar discs -- Modeling kinematics and morphologies}

   \author{Enrico Ragusa\inst{\ref{unimiMat},\ref{ENS},\ref{unimi}},
   Elliot Lynch\inst{\ref{ENS}},
   Guillaume Laibe\inst{\ref{ENS}},
   Cristiano Longarini\inst{\ref{unimi},\ref{ioa}},
   Simone Ceppi\inst{\ref{unimi}}
          }
   \authorrunning{E. Ragusa et al.}

   \institute{Dipartimento di Matematica, Università degli Studi di Milano, Via Saldini 50, 20133, Milano, Italy \label{unimiMat}
       \and
   ENS de Lyon, CRAL UMR5574, Universite Claude Bernard Lyon 1, CNRS, Lyon, F-69007, France\label{ENS}
        \and         
    Dipartimento di Fisica, Università degli Studi di Milano, Via Celoria 16, 20133 Milano MI, Italy\label{unimi}
    \and
    Institute of Astronomy, University of Cambridge, Madingley Road, Cambridge, CB3 0HA, United Kingdom\label{ioa}\\
   \email{enrico.ragusa@unimi.it}
             }

   \date{Received September 15, 1996; accepted March 16, 1997}
 
  \abstract
   {Protostellar discs are mostly modelled as circular structures of gas and dust orbiting a protostar. However, a number of physical mechanisms, e.g. the presence of a (sub)stellar companion or initial axial asymmetry, can cause the gas and dust orbital motion to become eccentric. Theoretical studies have revealed that, when present, disc eccentricity is expected to occur with predictable profiles that can be long-lasting and potentially observable in protostellar systems. 
   }
   {We  construct an analytical model predicting the typical features of the kinematics and morphology of eccentric protostellar discs, with the final goal of characterising the observational appearance of eccentricity in discs.}
   {We validate the model using a numerical simulation of a circumbinary disc (where the binary makes the disc eccentric). We finally post-process the simulation with Monte Carlo Radiative Transfer to study how eccentric features would appear through the ``eyes'' of ALMA. }
   {Besides the 
   motion of the material on eccentric Keplerian orbits in the disc orbital plane, the most characteristic eccentric feature emerging from the analytical model is strong vertical motion with a typical anti-symmetric pattern (with respect to the disc line of pericentres). A circumbinary disc with a $\approx 40$ au eccentric cavity ($e_{\rm cav}=0.2$), carved by an $a_{\rm bin}=15$ au binary, placed at a distance $d=130$ pc, is expected to host in its upper emission surface vertical oscillations up to $v_{z}\sim 400\, {\rm ms}^{-1}$ close to the cavity edge, i.e. well within ALMA spectral and spatial resolution capabilities.
   A residual spiral pattern in the vertical velocity $\Delta v_{z}\sim 150\, {\rm ms}^{-1}$ of the simulation cannot be captured by the theoretical model, we speculate it to be possibly linked to the presence of a companion in the system. 
}
   {}


   \maketitle
%

\section{Introduction}
In celestial mechanics, particles orbiting spherically symmetric objects follow Keplerian ellipses. This classical physics result, at the heart of solar and extrasolar planet mechanics for centuries, comes from Newton's $1/r$ potential. These orbits naturally extend to continuous distributions of matter orbiting point-like masses, known as accretion discs, where the horizontal fluid motion follows a set of nested confocal, Keplerian ellipses. 

The mathematical complexity of describing the evolution of fluid flows in eccentric discs has meant that circular discs are almost always preferred as ``spherical cows'' to derive models that can be compared with observations  (e.g. \citealp{shakura1973,pringle1981}), filtering out effects such as variable scale heights and non-axisymmetric orbital compression.  In that sense, the widely used terminology ``Keplerian discs'', abusively used for ``circular Keplerian discs'', is misleading, since it propagates the idea that discs have no eccentricity by default.

The dynamics of eccentric discs has been the subject of a number of theoretical studies (e.g. \citealp{ogilvie2001,goodchild&ogilvie2006,ogilvie2014,teyssandier2016,ogilvielynch2019,lee2019a,lee2019b}), which have shown that the eccentricity propagates in the disc as a form of density waves. In these formalisms the (generally nonlinear) radial and azimuthal velocity perturbations from circular motion can be expressed in terms of orbital eccentricity and the orientation of the apsidal angle. Disc eccentricity can be excited by a companion in the disc (binary star, planet, e.g. \citealp{lubow1991a,kley2006}). It has also been shown to be damped by bulk viscosity, but is generally excited by shear viscosity (e.g. \citealp{syer1992,kley1993}). 

However, it is not clear that viscous models of turbulence are even applicable to eccentric discs, making the evolution of eccentric discs a promising approach to constraining the behaviour of real disc turbulence -- see e.g. \citet{ogilvie2001,ogilvie2003b,ogilvie2003c,lynch2021} for theoretical works, and e.g.  \citet{papaloizou2005,pierens2020,dewberry2020,oyang2021,chan2022} for numerical works concerning turbulence in eccentric discs. 

Disc eccentricity also place a chronological constraint on scenarios of planet formation, with recent work showing that discs are likely born eccentric (\citealp{hennebelle2020,lebreuilly2021}; Lovascio et al. submitted), consistent with the young non-axisymmetric discs observed in the eDisk survey \citep{ohashi2023,thieme2023,vantoff2023}. This contrasts with the large fraction of more evolved discs that are observed to be nearly circular \citep{long2018,andrews2018,oberg2018}, with some exceptions represented by some discs with central cavities (often referred to as transition discs) where kinematic or morphological measurements of protostellar disc eccentricity were possible \citep{dong2018,kuo2022,garg2022,yang2023,kozdon2023}; other systems showed non-axisymmetric residuals when a circular Keplerian velocity map is subtracted from their kinematics \citep{wolfer2023}, possibly hinting at their eccentric nature. The future availability of new high spatial and spectral resolution datasets (e.g. the coming exoALMA survey) creates the need for new detailed models of disc kinematics and enables also the possibility of testing them directly with observations. 

Hence, in this study, we aim to characterise kinematic signatures of eccentricity in protostellar accretion discs. We show how the kinematics and morphology of eccentric discs is fully described by fixing three functions: the eccentricity profile $e(a)$, the pericentre longitude profile $\varpi(a)$, the disc mass distribution $M_a(a)$, acting as a sort of density profile that is constant on orbits, and an assumption about the disc thermodynamics, here done prescribing a locally isothermal sound speed profile $\langle \cs \rangle(a)$. In the definition of the profiles we just discussed, $a$ is the disc semi-major axis coordinate from celestial mechanics. 

We finally note, for completeness, that eccentric disc dynamics has been proven to be relevant in other astrophysical contexts than protostellar discs: e.g., disc eccentricity has been directly measured and discussed in debris discs in the late stages of star formation \citep{macgregor2013,olofsson2019,faramaz2019,booth2021,macgregor2022,lynch2022,lovell2023}; similarly debris discs formed after the tidal disruption of asteroids around white dwarfs feature changes in the shape of spectral lines that have been interpreted as the precession of an eccentric disc due to pressure and/or to general relativistic effects \citep{wilson2015,manser2016,manser2016b,cauley2018,miranda2018,dennihy2018,dennihy2020}; finally, in tidal disruption events of stars around black holes where the disc formed after the stellar disruption is expected to be eccentric \citep{liu2017,cao2018,hung2020,wevers2022}. 

This paper is divided as follows: in Sec. \ref{sec:theorkinstruct}, we provide a simplified analytical model that, taking as input $e(a)$, $\varpi(a)$, $M_a(a)$ and $\langle \cs \rangle(a)$, predicts: the radial motion $v_R$, the orbital motion $v_\theta$, the vertical motion $v_z$ and disc morphology, in the form of $H/R$ and $\Sigma(a,\phi)$ -- i.e. predicting over-dense regions due to the eccentric nature of the disc. We compare the eccentric disc theoretical model developed to a numerical simulation of an eccentric circumbinary disc in Sec. \ref{sec:simulations} and \ref{sec:discussion}. Finally, in Sec. \ref{sec:MCRT}, we examine the observability of these features in protostellar discs using ALMA. To do so, we rescale the system and perform Monte Carlo radiative transfer simulations to quantify the amplitude of the velocity perturbations. We draw our conclusions in Sec. \ref{sec:conclusion}.



\section{Analytic representation of the disc kinematics and structure}\label{sec:theorkinstruct}

\begin{figure}
   \centering
   \includegraphics[width=\columnwidth]{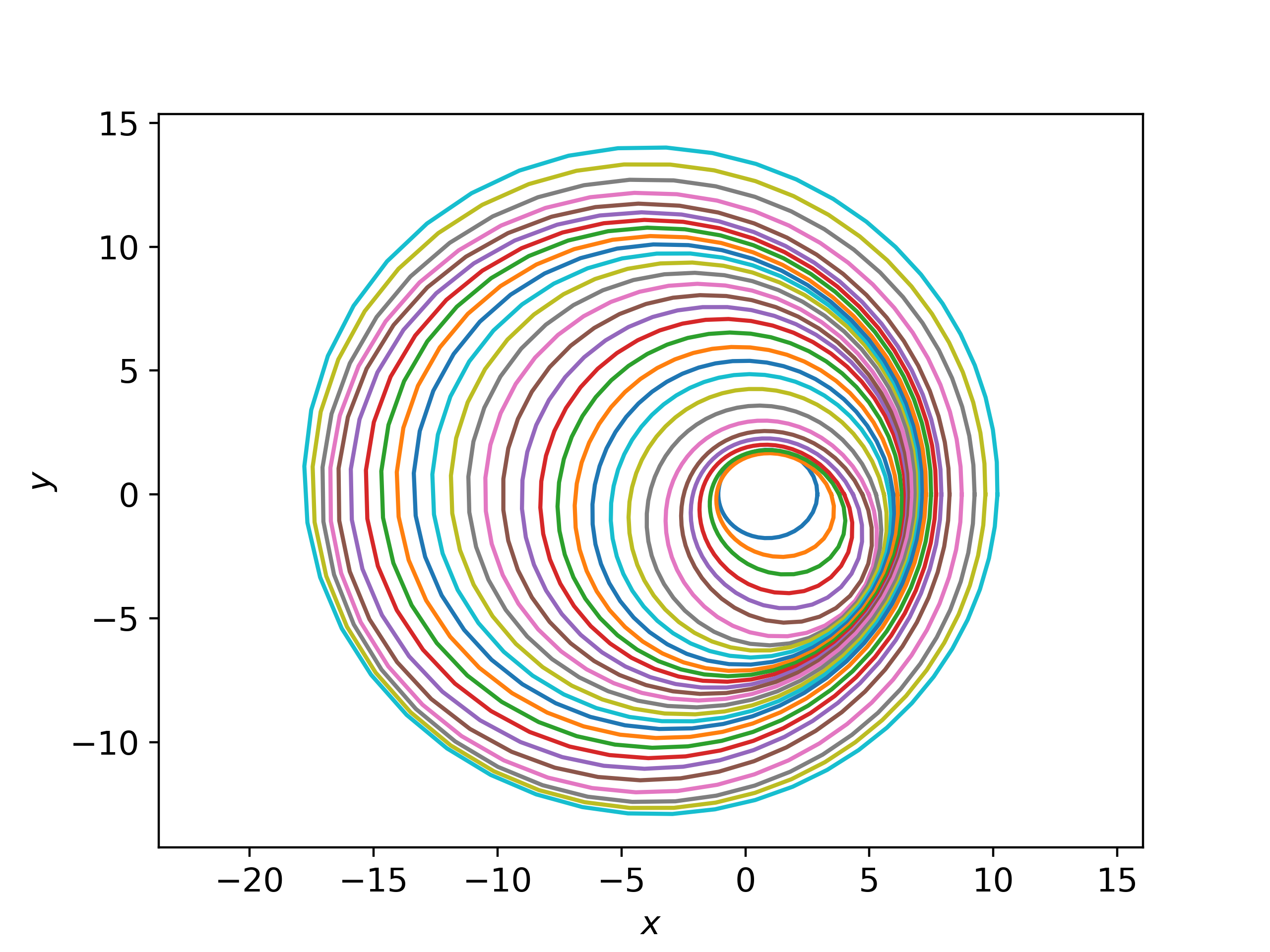}
   \caption{Example of the morphology of a disc composed by a set of nested ellipses with both eccentricity $e(a)$ and pericentre phase $\varpi(a)$ varying for different values of the semi-major axis.}
              \label{fig:nested}%
\end{figure}

Eccentric discs can be described as a continuous set of nested, confocal, ellipses, orbiting the baricentre located at the common focus, with an eccentricity $e(a)$ and pericentre phase $\varpi(a)$ profiles (see Fig. \ref{fig:nested}), where $a$ is the semi-major axis of the ellipse. Below we summarise some key aspects concerning the formalism used for describing their structure and dynamics. Details concerning eccentric disc evolution (i.e. how the eccentricity profile evolves) are beyond the scope of the paper, however we refer the reader to Appendix \ref{appendix:moreinfo} and references therein for a quick overview of the theory. 

To do so, as is common in eccentric disc literature (e.g. \citealp{ogilvie2001,ogilvie2014}), we introduce in Sec. \ref{sec:ecccoord} a new coordinate system specialised to deal with eccentric orbits, then putting together celestial mechanics results, hydrodynamic considerations we provide a description of eccentric discs morphology and kinematics.
The model presented summarises results from the existing literature, here self-consistently re-derived to unify the notation, for an $(a,\phi,z)$ coordinate system in Eulerian form in Sec. \ref{sec:ecccoord}, \ref{sec:orbvel}, \ref{sec:vertmotion}, \ref{sec:morphology}. We also introduce a new effective prescription for pressure corrections of the azimuthal velocity to simplify the equations in \ref{sec:presscorr}.

\subsection{Eccentric discs coordinates}\label{sec:ecccoord}
In this section we introduce a convenient coordinate system for describing the kinematics and structure of eccentric discs.

We first note that using cylindrical coordinates $(R,\theta,z)$ when dealing with eccentric disc geometry is very inconvenient: a single value of $R$ spans through multiple elliptical orbits for different values of $\theta$. As mentioned above, since the orbital structure of an eccentric disc is defined by $e(a)$ and $\varpi(a)$, a much better choice is an elliptical coordinate system using $(a,\phi)$, where $a$ is the semi-major axis of the ellipse, while the azimuthal coordinate $\phi$ is related to the true anomaly by $f=\phi-\varpi(a)$. Since we assume that the focus of the ellipses is fixed at the origin, $\phi$ is coincident with the polar azimuthal coordinate $\theta$. 

Assuming that the focus of all the confocal ellipses is at the origin of the cartesian coordinate system $(x,y,z)$, the transformation rules between $(a,\phi,z)\rightarrow (x,y,z)$ coordinates are
\begin{align}
x(a,\phi)&=R(a,\phi)\cos[\phi]\label{eq:xyaf1}\\
y(a,\phi)&=R(a,\phi)\sin[\phi]\label{eq:xyaf2},\\
z=z,
\end{align}
where $R(a,\phi)$ is
\begin{equation}
R(a,\phi)=\frac{a[1-e(a)^2]}{1+e(a)\cos[\phi-\varpi(a)]}.\label{eq:Raf}
\end{equation}
The Jacobian matrix from this transformation is
\begin{equation}
\mathcal{J}=\frac{\partial(x,y,z)}{\partial(a,\phi,z)}.
\end{equation}
Its determinant, $J(a,\phi)=\det(\mathcal{J})$, will be useful for a number of purposes below, it has the dimension of a length and and it is given by
\begin{align}
    J(a,\phi)&=R(a,\phi)^3\frac{1-e(a)[e(a)+2ae_a(a)]}{a^2[1-e(a)^2]^2}\nonumber\\
    &\quad\quad\quad\quad\quad\quad \cdot\left\{1-q\cos[\phi-\varpi(a)-\alpha]\right\},\label{eq:J}
\end{align}
where the angle $\alpha$ and variable $q$ are introduced to simplify the form of the equations, by exploiting the relation for $\cos(a-b)$, and are related to $e$, $a e_a$, and $a\varpi_a$ by
\begin{align}
q\cos(\alpha)&=\frac{[1 + e(a)^2] ae_a(a)-[1 - e(a)^2]e(a)}{1-e(a)[e(a)+2ae_a(a)]},\label{eq:qcosa}\\
q\sin(\alpha)&=\frac{e(a)[1-e(a)^2]a\varpi_a(a)}{1-e(a)[e(a)+2ae_a(a)]}.\label{eq:qsina}
\end{align}
The quantities $e_a(a)$ and $\varpi_a(a)$ indicate the derivative with respect to $a$ of $e(a)$ and $\varpi(a)$, and are often referred to as ``eccentricity gradient'' and ``twist'', respectively. 
Eq. (\ref{eq:qcosa}) and (\ref{eq:qsina}) also imply that the angle $\alpha$ in Eq. (\ref{eq:J}) satisfies
\begin{equation}  \tan(\alpha)=\frac{ae(a)[1-e(a)^2]\varpi_a(a)}{[1 + e(a)^2] ae_a(a)-[1 - e(a)^2]e(a)}.\label{eq:alpha}
\end{equation}

We note that since the coordinate $z$ does not depend on $a$ and $\phi$, the expression of $J$ is the same both when we deal with a 2D disc using the $(a,\phi)$ coordinates only, and for a full 3D disc $(a,\phi,z)$.

In order to provide a physical interpretation to $J$, we note that it constitutes the multiplicative factor to preserve element volume ${\rm d}V=\diff x \diff y \diff z$ or area $\diff A=\diff x\diff y$ while transforming to eccentric coordinates as follows
\begin{align}
    \diff V&=J\diff a\diff \phi \diff z,\\
    \diff A&=J\diff a\diff \phi.
\end{align}

In the light of this, we note that a vanishing value of $J$ indicates a coordinate singularity (one value of $a$ characterises multiple ellipses at that location), the nested ellipses overlap causing an orbit intersection (e.g. \citealp{statler2001}). From Eq. (\ref{eq:J}), it is easy to verify that this instance occurs when $q\geq 1$.
Similarly $\alpha$ represents the true anomaly, i.e. the angle from the pericentre longitude $\varpi$, where the maximum orbital compression occurs.

\subsection{Orbital velocity field: Keplerian motion} \label{sec:orbvel}

In this section we define the velocities in the ``eccentric'' $(a,\phi)$ coordinate system.

From celestial mechanics we know that the radial and azimuthal orbital velocities, in cylindrical coordinates $(R,\theta)$, of a test mass in Keplerian motion around a central body are \citep{murray1999}
\begin{align}
v_{R}(a,\phi)&=  \dot R=  a\Omega_0\frac{e(a)}{\sqrt{1-e^2(a)}}\sin[\phi-\varpi(a)],\label{eq:vr}\\
v_{\theta,{\rm K}}(a,\phi)&=R\dot \theta=    a\Omega_0\frac{1+e(a) \cos[\phi-\varpi(a)]}{\sqrt{1-e^2(a)}},\label{eq:vphi}
\end{align}
where the dot notation indicates total time derivative, ${\rm d}/{\rm d}t$, and $\Omega_0$ is the mean motion
\begin{equation}
    \Omega_0=\sqrt{\frac{GM}{a^3}}.\label{eq:orbfreq}
\end{equation}
From Eq. (\ref{eq:vphi}) we define the angular orbital velocity along an eccentric orbit as
\begin{equation}
    \Omega(a,\phi)\equiv\dot \theta=\Omega_0\frac{\{1+e(a)\cos[\phi-\varpi(a)]\}^2}{[1-e^2(a)]^{3/2}}\quad.\label{eq:omega}
\end{equation}

The standard analytical treatment of differential operators in the eccentric coordinate system requires the definition of contravariant velocities as $v^a\equiv \dot a$ and $v^\phi\equiv\dot \phi$.
With these definitions in mind we can define $v^a$ and $v^\phi$ consistently with the orbital velocities in Eq. (\ref{eq:vr}) and (\ref{eq:vphi}). 
Along an eccentric orbit the semi-major axis does not change, so that 
\begin{equation}
     v^a=0,
\end{equation} 
while, given the coincidence between $\phi$ and $\theta$, $v^\phi$ is
\begin{equation}
    v^\phi=\Omega.
\end{equation}
We remark that $v^\phi$ is a contravariant variable, with the dimension of an angular velocity. While $v_\theta$ is the azimuthal velocity, with the dimension of a standard spatial velocity, not the covariant velocity component that might be expected from the notation. In this context, $v_\theta$ and $v^\phi$ have different dimensions and should not be confused. 

\subsection{Vertical motion}\label{sec:vertmotion}

During an eccentric orbit the material undergoes vertical oscillations driven by the periodic variation of the vertical gravitational potential along the orbit (the distance from the central source of gravity varies through the orbit \citealp{ogilvie2014}).
Since we use a new coordinate system, we write the continuity equation in contravariant form,
\begin{align}
    \pd{\rho}{t}&=-\frac{1}{J}\pd{}{a}(Jv^a \rho)-\frac{1}{J}\pd{}{\phi}(Jv^\phi\rho)-\pd{}{z}(v^z\rho) , \label{eq:continuityrho}
\end{align}
obtained from the expression of the divergence of a vector field $\bm u$ in a generic coordinate system $\{x^i\}$,
\begin{equation}    
\nabla_i(u^i)=\frac{1}{J}\frac{\partial}{\partial x_i}Ju^i.
\end{equation}
For our purpose we also need the equation for momentum conservation in the vertical direction, 
\begin{equation}
\pd{v^z}{t}+v^a\pd{v^z}{a}+v^\phi\pd{v^z}{\phi}+v^z\pd{v^z}{z}=-\frac{GM}{R^2}\frac{z}{R}-\frac{1}{\rho}\pd{p}{z}. \label{eq:vertvel1}
\end{equation}

We now look for steady solutions and substitute $v^a=0$ and $v^\phi=\Omega(a,\phi)$ -- as we prescribed in Sec. \ref{sec:orbvel}. After these operations, Eq. (\ref{eq:continuityrho}) and (\ref{eq:vertvel1}) reduce to
\begin{align}
 \pd{}{z}(v^z\rho)&=-\frac{1}{J}\pd{}{\phi}(J\Omega\rho),\label{eq:veloverteq21}\\
\Omega\pd{v^z}{\phi}+v^z\pd{v^z}{z}&=-\frac{GM}{R^2}\frac{z}{R}-\frac{1}{\rho}\pd{p}{z}.\label{eq:veloverteq22}
\end{align}

We assume the disc to be locally isothermal so that $\rho$ and $p$ have Gaussian form (e.g. \citealp{lodato2008} for a review of classical disc dynamics)
\begin{align}
\rho&=\frac{\rho_0}{\sqrt{2{\rm \pi}}}\exp\left(-\frac{z^2}{2H^2}\right),\label{eq:rho}\\
p&=\frac{p_0}{\rho_0}\rho,\label{eq:p}
\end{align}
where $p_0$ and $\rho_0$ represent the density value in the midplane. We note that $p_0/\rho_0=\cs^2$ is the squared local isothermal sound speed at the disc midplane; $H(a,\phi)$, is the standard deviation of the vertical density distribution, here representing the disc local vertical scale-height\footnote{In the general framework (e.g. when the disc is not locally isothermal), $H$ is defined by the second vertical moment of the density distribution as $ H^2=\int_z z^2\rho\diff z/\int_z\rho\diff z$.}. The form of the final equations does not change for different assumptions on $\rho$ and $p$, with some caveats discussed in Appendix \ref{appendix:separable}.

In order to study the evolution of $H$ along the orbit, we need a reasonable ansatz for $v^z$. We impose the quantity $z/H$ to be a Lagrangian variable -- i.e. $z/H$ does not change along the orbit -- that follows the disc expansion 
\begin{equation}
    \frac{\rm d}{{\rm d}t}\frac{z}{H}=\pd{}{t}\frac{z}{H}+\Omega\pd{}{\phi}\frac{z}{H}+v^z\pd{}{z}\frac{z}{H}=0.\label{eq:lagrzh}
\end{equation}
This ansatz is physically the homogeneous vertical expansion of the gas column.
When the disc is stationary ($\partial/\partial t=0$) such a condition is satisfied if $v^z$ has the form\footnote{By making explicit the azimuthal dependence of $\Omega=\Omega_0\omega(\phi)$ and $H=H_0h(\phi)$ it is possible to rewrite 
    \begin{equation}
    v^z\equiv \cs\left[\omega(\phi)\pd{h}{\phi}\right]\frac{z}{H}.
\end{equation}
where $\cs=H_0\Omega_0$, highlighting that $\cs$ sets the natural scale for the velocity $v^z$.\label{fn:csnorm}}
\begin{equation}
    v^z=\Omega\pd{H}{\phi}\frac{z}{H}.\label{eq:vz}
\end{equation}
We use Eq. (\ref{eq:vz}) as an ansantz for the solution of $v^z$.
By substituting this into Eq. (\ref{eq:veloverteq21}) and (\ref{eq:veloverteq22}), they can be re-written as
\begin{align}
\Omega\pd{\rho_0}{\phi}&=-\rho_0\left[\frac{1}{J}\pd{}{\phi}(J\Omega)+\frac{\Omega}{H}\pd{H}{\phi}\right],\label{eq:veloverteq31}\\
\Omega^2\frac{\partial ^2 H}{\partial \phi^2}&=-\Omega\pd{H}{\phi}\pd{\Omega}{\phi}-\frac{GM}{R^3}H+\frac{\cs^2}{ H}.\label{eq:veloverteq32} 
\end{align}
A similar set of equations can be obtained from the Lagrangian viewpoint (e.g. \citealp{ogilvie2014,lynch2021}).

\begin{figure}
   \centering  \includegraphics[width=\columnwidth]{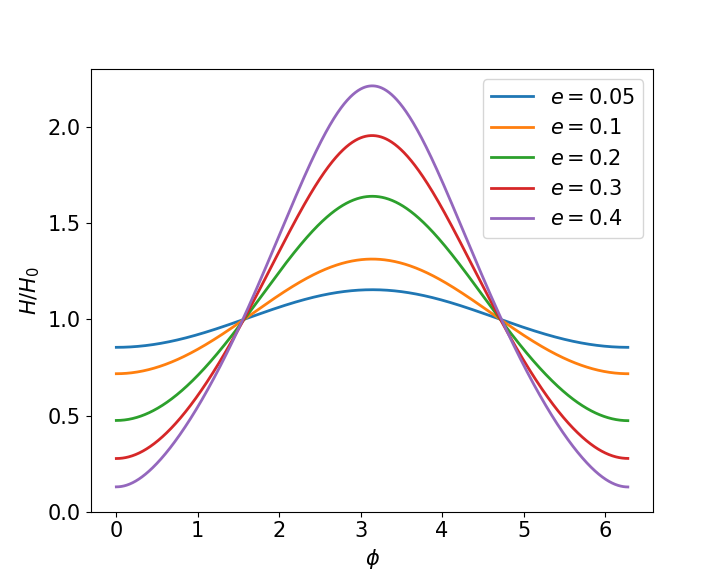}  \includegraphics[width=\columnwidth]{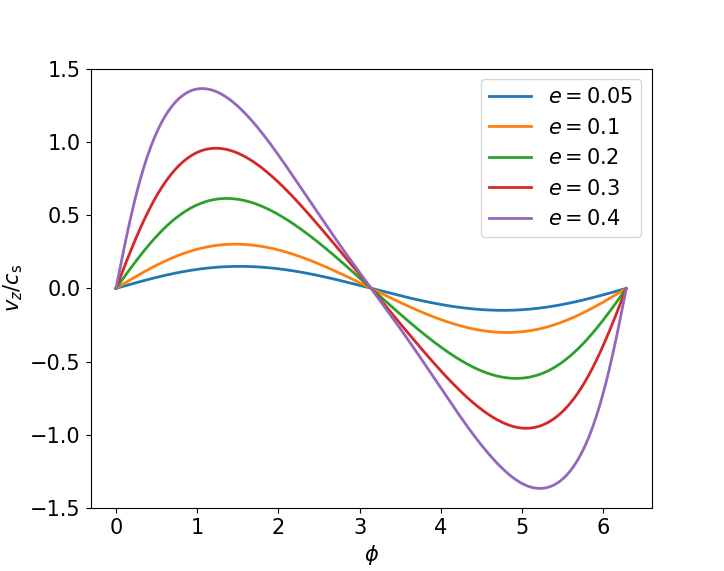}
\caption{Top panel: $H/H_0$ as a function of $\phi$, where $H_0=\cs/\Omega_0$, obtained solving Eq. (\ref{eq:veloverteq32}). Bottom panel: $v_z/\cs$ as a function of $\phi$, obtained from Eq. (\ref{eq:vz}) after solving Eq. (\ref{eq:veloverteq32}), calculated at an elevation $z=H$ above the midplane -- $\cs$ is the most natural normalisation$^{\ref{fn:csnorm}}$ for $v_z$. Solution to Eq. (\ref{eq:veloverteq32}) was obtained solving for periodic solutions across one orbit, i.e. $H(a,0)=H(a,2{\rm \pi})$, using a shooting method for solving the boundary value problem.}
   \label{fig:Hvsphi}
\end{figure}

We first solve Eq. (\ref{eq:veloverteq32}), as it does not depend on Eq. (\ref{eq:veloverteq31}). In particular, Fig. \ref{fig:Hvsphi} shows the solutions to Eq. (\ref{eq:veloverteq32}) for the azimuthal dependence of $H$ for different values of eccentricity and fixed $\varpi=0$, and the resulting $v^z$ from Eq. (\ref{eq:vz}).
The value of $\rho_0$ solving Eq. (\ref{eq:veloverteq31}) can then be easily obtained by noting that Eq. (\ref{eq:veloverteq31}) can be rewritten as
\begin{equation}
    \frac{1}{J}\pd{}{\phi}(J\Omega H\rho_0)=0,\label{eq:cons3d}
\end{equation}
implying that $J\Omega H\rho_0\equiv \mathcal F(a)$ depends on the semi-major axis only and is a conserved mass flux along the orbit. Further considerations on $\mathcal F(a)$ will be used in the next section to discuss the disc morphology.

To conclude this section, we comment on the validity of the assumption that the disc is locally isothermal. This assumption implies that the gas temperature does not experience changes in its temperature due to compression or expansion, even though for large eccentricities the disc might undergo a very strong vertical expansion/compression at the apocentre/pericentre. Relaxing the assumption of the disc being locally isothermal implies treating also the internal energy or assuming an adiabatic equation of state. This would cause pressure forces to increase more steeply during the compression at the pericentre, reducing the ratio between maximum and minimum disc scale-height at apocentre and pericentre $H_{\rm apo}/H_{\rm peri}$. A full derivation with an adiabatic equation of state is beyond the scope of the paper, but can be found in \citet{ogilvie2014} or \citet{lynch2021}. 

Beyond values of eccentricity of $e\approx 0.5\textrm{--}0.6$ the value of the $\cs/H$ term in \ref{eq:veloverteq32} can become very large close to pericentre, causing integration issues. 

\subsection{Morphology}\label{sec:morphology}

Eccentric discs might present steady over-densities, caused by orbital compression (i.e. $e_a\neq 0$ and $\varpi_a\neq 0$), that result in a characteristic non-axisymmetric morphology.
In this section we put together the previous geometrical considerations and conservation of flux along the orbit to derive the expression of $\Sigma(a,\phi)$. 

We obtained in Eq. (\ref{eq:cons3d}) conservation of mass flux along elliptical orbits. We note that $\Sigma=\int_z \rho {\rm d}z=H\rho_0$, thus we can rewrite $\mathcal F (a)$ as\footnote{Eq. (\ref{eq:flux}) can be directly obtained in this form by vertically integrating the continuity equation Eq. (\ref{eq:continuityrho}) using $\Sigma=\int_z \rho \diff z$ and looking for stationary solutions.}
\begin{equation}
 \mathcal F(a)= J(a,\phi) \Omega(a,\phi) \Sigma(a,\phi). \label{eq:flux}
\end{equation}

We note that integrating the r.h.s. of Eq. (\ref{eq:flux}) over an orbital time $t_{\rm orb}$ implies
\begin{equation}
    \int_0^{t_{\rm orb}}J(a,\phi)\Sigma(a,\phi)\Omega(a,\phi)\diff t=\int_0^{2{\rm \pi}}J(a,\phi) \Sigma(a,\phi) \diff \phi= M_a(a)
\end{equation}
where $M_a(a)$ is the derivative of the total disc mass enclosed within the semi-major axis $a$, $M(a)$, with respect to the semi-major axis -- i.e. $M_a\diff a$ is the mass of an infinitesimal eccentric annulus of width $\diff a$. 

Since $\mathcal F(a)$ does not depend on $\phi$, we have $t_{\rm orb} \mathcal F(a)=M_a(a)$, which can be rewritten as $\mathcal F(a)=M_a(a)\Omega_0/2{\rm \pi}$, where we recall $\Omega_{0}$ is the mean motion (Eq. \ref{eq:orbfreq}).
In the light of these considerations, Eq. (\ref{eq:flux}) can be rewritten as \citep{ogilvie2001,statler2001}
\begin{equation}
    \Sigma(a,\phi)=\frac{M_a(a)\Omega_{0}}{2{\rm \pi}J(a,\phi)\Omega(a,\phi)}.\label{eq:sigmamod}
\end{equation}

If $e_a(a)=0$ and $\varpi_a(a)=0$ -- i.e., the disc has vanishing eccentricity gradient and twist -- the expression of $\Sigma(a,\phi)$ simplifies to an expression that does not depend on $\phi$. 

This implies that an eccentric disc can, in principle, show no morphological features at all for arbitrary large eccentricities, provided its eccentricity and pericentre phase profiles are constant. Conversely, a negative (positive) eccentricity gradient $e_a(a)<0$ ($e_a(a)>0$) will produce an azimuthal overdensity located at the apocentre (pericentre); if the disc also presents some orbit twisting ($\varpi_a(a)\neq 0$), the overdensity is located at angle $\phi$ satisfying $\phi-\varpi(a)=\alpha$, where $\alpha$ is presented in Eq. (\ref{eq:alpha}).

It is important to note that any $e(a)={\rm const}$, $\varpi(a)={\rm const}$ profile is not steady. Such a disc will evolve in time producing a strong eccentricity gradient and twist, due to the transport throughout the disc of angular momentum and energy in the form of eccentric waves. If some dissipation is present, the disc will finally settle in a steady eccentricity configuration. More considerations concerning the evolution of eccentricity profiles can be found in Appendix \ref{appendix:eccdiscdyn}.

\subsection{Approximated pressure corrections}\label{sec:presscorr}
In a pressure supported disc the centrifugal balance is reached with a sub/super-Keplerian orbital velocity. 
In analogy with classic disc accretion theory, we apply a generalisation of the standard pressure correction $\delta v_{\theta,{\rm P}}$ required for a circular accretion disc, i.e. (e.g. \citealp{lodato2008} for a review of classical disc dynamics):
\begin{align}
    v_{\theta}&=v_{\theta,{\rm K}}\sqrt{1+\eta},\label{eq:vpress}\\
    \eta&=(\Omega_0 a)^{-2}\left(\frac{a}{\Sigma_0}\dd{P}{a}\right)\\
    P&=\langle\cs\rangle^2\Sigma_0,
\end{align}
where $v_{\theta,{\rm K}}$ is the Keplerian velocity (Eq. \ref{eq:vphi}), $\Sigma_0$ constitutes an averaged value along the orbit defined as
\begin{equation}
    \Sigma_0=\frac{M_a}{2{\rm \pi}a},
\end{equation}
and $\langle\cs\rangle$ is an averaged value of $\cs$ along an orbit with semi-major axis $a$.

Eq. (\ref{eq:vpress}) constitutes a simplification with respect to solving the fluid dynamics equations accounting for the pressure term. The analysis in Appendix A of \citet{ogilvielynch2019} allows for a self-consistent derivation of the pressure correction within the Hamiltonian hydrodynamics framework, however for the clarity of presentation we prefer to stick to the simplified approach provided in this section, that provides a good match with numerical simulations anyway, as will be discussed in Sec. \ref{sec:simulations}.

\subsection{Disc eccentricity evolution: growth and damping}

We summarise here for completeness the main physical mechanisms responsible for disc eccentricity growth and damping, despite remarking that the model we developed is independent of how the system arrived in its eccentricity configuration. More considerations concerning the existence and evolution of secularly evolving eccentric eigenmodes is discussed in Appendix \ref{appendix:eccdiscdyn}.

Concerning growth, two main mechanisms can be found: eccentricity pumping by a companion mass and primordial growth of the disc eccentricity. 

In the first, a second mass in the system\footnote{Binary stars or planets for protostellar systems -- but also moons, for planets with rings like Saturn, or compact objects, for X-ray binaries, cataclismic variables and stellar/supermassive binary black hole.}, above a certain mass-ratio threshold ($q\gtrsim 10^{-3}$, \citealp{kley2006}) injects angular momentum and energy at resonant locations that cause the disc eccentricity to grow \citep{lubow1991a,goldreich2003}. This has been confirmed by numerical studies explicitly investigating the evolution of the disc eccentricity
assuming both low mass-ratio companions \citep{papaloizou2001,kley2006,dangelo2006,dunhill2013,ragusa2018,teyssandier2019,dempsey2021,tanaka2022} and high mass-ratio companions, both in circumbinary discs \citep{macfadyen2008,marzari2009,shi2012,dunhill2015,miranda2017,ragusa2020,munoz2020,pierens2020,dittman2022,siwek2023} and in the discs surrounding the individual masses \citep{lubow1991a,lubow1991b,whitehurst1994,murray1996,kley2008,regaly2011}. Numerical simulations of flybys -- i.e. the two stars are not gravitationally bound -- in star formation regions have also measured a substantial growth of the disc eccentricity after the close encouter between the two stars \citep{cuello2019}. Furthermore numerous numerical works on circumbinary discs, despite not measuring directly the disc eccentricity, highlighted the formation of visibly eccentric cavities  (e.g. \citealp{dorazio2013,farris2014,ragusa2016,ragusa2017,price2018,calcino2019,calcino2020,heath2020,tiede2020,franchini2023}) or eccentric gaps for planets and low mass-ratio companions (e.g. \citealp{ataiee2013,zhu14b,scardoni2023}). 

A second relevant scenario for eccentric discs concerns the primordial deviations from circular motion of the material in the discs that were excited and impressed during its formation -- i.e. the disc was born eccentric. 

In the process of star formation, it is reasonable to expect that the phase of cloud collapse is not spherically symmetric, seeding an initial eccentricity in the protostellar disc. Numerical simulations studying this phase appear to produce, as a natural outcome, eccentric discs, even in presence of symmetric initial conditions (Lovascio et al. submitted). 
Similarly, tidal disruption events (TDE) of eccentric (or parabolic) bodies around compact objects are expected to form eccentric discs surrounding the central body -- be it a white dwarf \citep{tavascus2021}, neutron star \citep{kurban2023} or a black hole \citep{shiokawa2015,bonnerot2016,zanazzi2020,lynch2021,cufari2022}. In this scenario, the debris are expected to inherit the eccentricity of the disrupted body -- in the event of parabolic orbits as in the context of TDEs of stars on supermassive black holes, half of the disrupted body remains bound to the central object with an eccentricity $e\lesssim 1$. 

Concerning disc eccentricity damping, we can identify a resonant and a viscous/dissipative mechanisms. On the one hand, while some resonances can pump the disc eccentricity, others can damp it \citep{goldreich2003}. Both eccentricity damping and pumping resonances can be present in a disc. Whether the disc eccentricity is growing or decreasing depends on the balance of the contributions of individual resonances, akin to what determines the direction of satellite migration: when some resonances push the satellite inward and others outwards. 

On the other hand, viscous/dissipative processes decrease the disc eccentricity. Viscosity is often included in theoretical models to describe turbulence, that, among its effects, causes the material to accrete on to the central mass. Pure shear, as for the \citet{shakura1973} $\alpha$-prescription, is not expected to produce eccentricity damping. In contrast, in some instances it can even trigger a viscous-overstability (e.g. \citealp{syer1992,kley1993,lyubarskij1994,latter2006}) that results in a growth of the disc eccentricity. Bulk viscosity instead always produce eccentricity damping \citep{ogilvie2001,goodchild&ogilvie2006,lynch2022b}.

Numerical dissipation, intrinsically present in any numerical scheme, produces a spurious form of bulk viscosity that causes eccentricity damping.
In general, this implies that any numerical simulation is affected by a spurious decrease of the disc eccentricity due to unavoidable numerical dissipation. This produces a damping rate that is most likely to be higher than what would be reasonable to expect if the fluid was inviscid and turbulent. To date, no reliable estimate about eccentricity damping rates is available. A study about the interplay between disc eccentricity and turbulence has been attempted by \citet{wienkers2018}. 

\subsection{How does the disc eccentric nature affect the dynamics}

In the previous sections we provided equations that describe the kinematics and structure of eccentric discs.
As obvious from celestial mechanics considerations, compared with the circular case where $v_{\theta,{\rm circ}}={\rm const}$ and $v_R=0$ along an orbit, an eccentric disc has $v_\theta$ and $v_R$ varying around the orbit. The azimuthal velocity $v_{\theta}$ (Eq. \ref{eq:vphi}, \ref{eq:vpress} for the pressure corrected) reaches its maximum at the phase of pericentre $\varpi$ and minimum at the phase of the orbit apocentre. Similarly, $v_R$ (Eq. \ref{eq:vr}) oscillates around $v_R=0$, reaching its absoulute maxima ($|v_R|$) at true anomaly $f={\rm \pi}/2$ and $f=3{\rm \pi}/2$. 

However, eccentric discs are also characterised by motion out of the disc plane, in the vertical direction.
This motion is the result of the change of the vertical gravitational field along the orbit that produces a change in the vertical scale height $H$ of the disc (Eq. \ref{eq:veloverteq32}). This effect causes $H$ to reach its minimum at pericentre and maximum at apocentre. The ratio between $H_{\rm apo}$ at apocentre and $H_{\rm peri}$ at pericentre depends on the orbital eccentricity and can easily reach $H_{\rm apo}/H_{\rm peri}\gtrsim 6$ for $e\gtrsim 0.4$ (see Fig. \ref{fig:Hvsphi}). It is useful to note that the ratio $z/H$ remains constant along the orbit (Eq. \ref{eq:lagrzh}), implying that the material moves along the orbit following eccentric streamlines with $z\propto H$.

This change of altitude along the orbit implies that the material develops a vertical velocity $v_z$ (Eq. \ref{eq:vz}) that oscillates between $v_z=0$ at the pericentre and apocentre and reaches its maximum absolute value $|v_z|$ at $f\sim {\rm \pi}/2$ and $f\sim 3{\rm \pi}/2$. 

The amplitude of these vertical $v_z$ oscillations scales with $z/H$ (Eq. \ref{eq:vz}). This implies that the oscillations increase in amplitude the higher a fluid element is in the disc atmosphere, and are anti-symmetric with respect to the midplane -- i.e. top and bottom layers of the disc have opposite sign of $v_z$. In particular, at $z=H$, the amplitude of the vertical oscillations has as natural scale factor the sound speed $\cs$ (see footnote \ref{fn:csnorm}). Given the linear scaling with $z$, $v_z$ can easily become supersonic in higher disc layers with respect to the midplane ($z>H$). This type of oscillations are often referred to as a ``breathing mode''.

Changes in the disc geometry due to eccentricity and pericentre phase variations across the disc result in overdensities and depletions of the disc surface density, $\Sigma$ (Eq. \ref{eq:sigmamod}). However, one should always keep in mind that a change in the disc thickness might affect the volume density, $\rho$, at the midplane without affecting the surface density: e.g. a disc with $e={\rm const}$ and $\varpi={\rm const}$ across its entire domain, produces no variations in the surface density morphology, i.e. $\Sigma(\theta)={\rm const}$. But, since the volume density at the midplane is $\rho=\Sigma/H$, $\rho$ will change as the inverse of $H$ -- i.e. $\rho$ will be maximum at pericentre and minimum at apocentre even when $\Sigma$ is constant along the orbit. 

\section{Numerical simulation}\label{sec:simulations}
In this section, we present a 3D hydrodynamical simulation of a circumbinary disc that throughout its evolution becomes eccentric (as expected due to binary-disc interaction) and perform a comparison with the analytical model predictions.
We note that for the remainder of the paper we will not use contravariant notation -- from now on we will refer to cylindrical (non-covariant) velocities as $v_R,\, v_\theta,\, v_z$ and to cartesian velocities as $v_x,\,v_y,\, v_z$.

We approach the problem as follows (Sec. \ref{sec:constr3d} for more details): we first extract from the simulations the profiles of $e(a)$, $\varpi(a)$ and $M(a)$, on which the analytical model depends. From these profiles, using the formalism discussed in Sec. \ref{sec:theorkinstruct}, we generate a 3D analytical model attributing to a grid of pixels $(x,y,z)$, representing the coordinates of the disc surface, the corresponding $(v_x,v_y,v_z)$. We finally compare the predictions of the theoretical model with the simulations.

We note that quantities generated using the theoretical model will be indicated with a subscript ``th'' (e.g. $x_{\rm th}$) while those calculated directly from the simulation with subscript ``sim'' (e.g. $x_{\rm sim}$).

\subsection{SPH numerical setup}\label{sec:hydrosph}

As a benchmark simulation to test the analytical model presented in Sec. \ref{sec:theorkinstruct} we use one of the simulations from the numerical set presented in \citet{ragusa2020}, i.e. a 3D hydrodynamical disc surrounding a binary (here used as a source for the disc eccentricity). 

Simulations in \citet{ragusa2020} were meant to study the evolution of the disc eccentricity exploring various choices of binary mass-ratios and disc parameters. All simulations with $M_2/M_1>0.05$ presented in \citet{ragusa2020} show a rapid growth of the disc eccentricity after $t\approx 400\textrm{--}500 t_{\rm bin}$ due to binary-disc interaction, and constitute the perfect testbed for the analytical model developed in this paper. 

In general, we note that the dynamics in the simulation is a priori richer than the one described by the theoretical model: viscosity, waves, quadrupolar potential, accretion and disc spreading are not at all considered in the derivation we provided above. 

\begin{figure}
   \centering
   \includegraphics[width=\columnwidth]{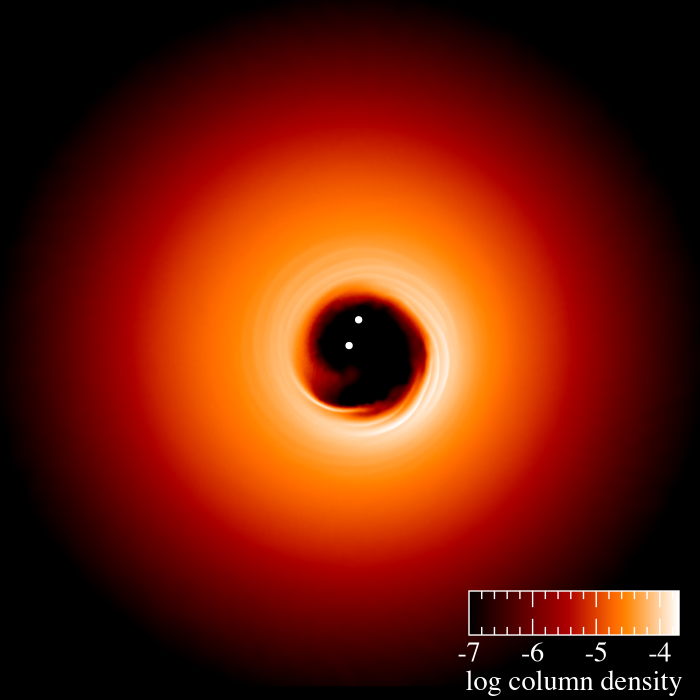}
   \caption{Rendering of the surface density of the dump used as reference numerical simulation for testing the analytical models. Dump from simulation 4A (mass-ratio $M_2/M_1=0.1$, $H/R=0.05$) in \citet{ragusa2020} after $t=500 t_{\rm bin}$. The dump features a cavity with an eccentricity of $e_{\rm cav}\approx 0.2$ and a cavity with a semi-major axis $a_{\rm cav}\approx 2.7 a_{\rm bin}$. }
\label{fig:sim}%
\end{figure}

From that set, we select simulation 4A (mass-ratio $M_2/M_1=0.1$). In particular, its $t=500 t_{\rm bin}$ dump (shown in Fig. \ref{fig:sim}), where $t_{\rm bin}$ is the binary orbital timescale represents a good candidate for our direct comparison: i) because at that time the disc eccentricity has reached a saturation value of $e_{\rm cav}\approx0.2$ at the cavity edge, which is qualitatively consistent with the few observationally measured ones in transition discs (e.g. \citealp{garg2022,yang2023}); ii) because the simulations with binary mass-ratios $M_2/M_1>0.1$ feature prominent density spirals and an orbiting overdense lump of material that alter the kinematics. The motion of the material in those simulations is in fact eccentric, but it is also characterised by additional perturbations that are not captured by the model in Sec. \ref{sec:theorkinstruct}. 

Simulation 4A of \citet{ragusa2020} was performed using the code \textsc{phantom} (SPH, \citealp{price2018a}), it used $N_p=10^6$ particles, and consisted of a circular, live binary with mass-ratio $M_2/M_1=0.1$ surrounded by a circumbinary disc with initial mass $M_{\rm d}= 5\cdot 10^{-3} \,M_{\rm bin}$ ($M_{\rm d}=4.8\cdot 10^{-3}\,M_{\rm bin}$ after $t=500\, t_{\rm bin}$), where $M_{\rm bin}=M_1+M_2$. The equation of state was chosen to be locally isothermal -- with radial temperature profile producing $\cs\propto R^{-0.25}$ and $H/R=0.05$ at $R_{\rm in}=2 a_{\rm bin}$. A circular cavity with $R=R_{\rm in}$ was initially excised at $t=0$. The mechanisms responsible for angular momentum transport were modeled using SPH artificial viscosity, producing an effective \citet{shakura1973} $\alpha_{\rm ss}\approx 5\cdot 10^{-3}$ (obtained using $\alpha_{\rm AV}=0.2$, $\beta=2$ \citealp{price2018a})\footnote{Such an implementation implies an unavoidable amount of bulk viscosity \citep{lodato2010}, in contrast with the pure shear nature of the \citet{shakura1973} prescription.} at the beginning of the simulation. 

\subsection{Obtaining maps of $v_R$, $v_\theta$, $v_z$, $H$, $\Sigma$ from the simulation }

For the comparison with the analytical model, we start generating projected $x-y$ maps of $v_{x,{\rm sim}}$, $v_{y,{\rm sim}}$, $v_{z,{\rm sim}}$, $H_{\rm sim}$ and  $\Sigma_{\rm sim}$ using vertical particle integration and density weighted integration -- i.e. an SPH equivalent of a density weighted average along the vertical direction. The calculation of these quantities is performed using \textsc{splash} as detailed below, the implementation of the integration procedures is described in \citet{price07a}. 

Velocities in the disc orbital plane are defined as $v_{x,{\rm sim}}=\langle v_{x}\rangle_z$ and  $v_{y,{\rm sim}}=\langle v_{y}\rangle_z$, where $\langle\cdot\rangle_z$ indicates density weighted integration along the vertical direction and $v_x$ and $v_y$ the velocities of individual SPH particles in the density weighted integration. Then, $v_{R,{\rm sim}}$ and $v_{\theta,{\rm sim}}$ are obtained projecting $\bm v_{\rm 2D}=\{v_{x,{\rm sim}},v_{y,{\rm sim}}\}$ on radial and azimuthal unit vectors, respectively. 

The map for $H_{\rm sim}$ is obtained through a density weighted integration along the vertical direction as $H_{\rm sim}=\langle z^2\rangle_z^{1/2}$ -- i.e., based on the definition of scale-height $H_{\rm sim}$ as the second moment of the particles vertical density distribution. Again, here $z^2$ indicates that the value of $z^2$ of individual SPH particles has been used in the density weighted integration. Since the disc is locally isothermal with a sound speed $\cs(R)$, the vertical density profile has the form of a Gaussian with standard-deviation $\sigma=H_{\rm sim}$.

Similarly, the map for $v_{z,{\rm sim}}$ is obtained performing a density weighted integration of the quantity $v_{z,{\rm sim}}=\sqrt{{\rm \pi}/2}\langle v_{z}{\rm sign}(z)\rangle_z$ along the vertical direction. This choice regarding the functional form and normalising pre-factor of the quantity to be vertically averaged has the following motivations: i) concerning the functional form, since $v_{z}$ is anti-symmetric with respect to the midplane, the quantity $v_{z}{\rm sign}(z)$ has the nice property of providing a $v_{z,{\rm sim}}$ maintaining the sign of the velocities of the $z>0$ half of the disc during the average across the midplane. In contrast, $\langle |v_{z}|\rangle_z$ or $\langle v_{z}^2\rangle_z$ would produce a map that is everywhere positive, while the vertical average of $\langle v_{z}\rangle_z$, would produce $v_{z,{\rm sim}}=0$ across the whole disc. ii) The pre-factor $\sqrt{{\rm \pi}/2}$ is introduced because, since the vertical density distribution is Gaussian, $\langle z{\rm sign} (z)\rangle_z= H\sqrt{2/{\rm \pi}}$. Under the assumption that $v_z\propto z/H$, as derived in Eq. (\ref{eq:vz}), the introduction of the pre-factor makes sure that $v_{z,{\rm sim}}$ traces the vertical velocity of particles at $z\sim H_{\rm sim}$ above the midplane. We anticipate that the results presented in the following sections confirm the validity of this assumption. 

Finally, the map of $\Sigma$ is obtained using the standard vertical integration of the particles' masses performed by \textsc{splash}. 

\subsection{Constructing a 3D analytical model from the hydrodynamical simulation}\label{sec:constr3d}

The entire set of equations constituting our analytical model to predict the kinematics and the morphology of eccentric discs rely exclusively on the profiles $e(a)$, $\varpi(a)$, their derivatives, $e_a$ and $\varpi_a$, $M_a$ and an assumption on the sound speed $c_s$, that here we choose to have a radial dependence $\cs(R)$ for consistency with what is prescribed in the simulation. In this section, starting from the aforementioned profiles, we generate a 3D analytical model attributing a velocity vector $(v_x,v_y,v_z)$ to each $(x,y,z)$, where $z$ represents the reference altitude of the disc surface -- this will be further clarified below.

From the reference snapshot discussed in Sec. \ref{sec:hydrosph} we extract the profiles of $e(a)$, $\varpi(a)$ and $M_a(a)$ (details about how semi-major axis profiles are calculated can be found in \citealp{teyssandier2017} and \citealp{ragusa2020}), we filter each profile using the Savitzky–Golay filter \citep{savitzky1964} to reduce the noise of the datasets and we interpolate them using cubic spline interpolation so that we can compute the derivatives $e_a(a)$, $\varpi_a(a)$ with respect to the semi-major axis. We set the filter window to be $w_{\rm sg}=50$ profile points (corresponding to a $\Delta a\sim 1.6a_{\rm bin}$) and polynomial order $p_{\rm sg}=2$. We note that filtering is a very important step, since noisy datasets of $e(a)$, $\varpi(a)$ would produce unusable outputs when taking derivatives with respect to $a$. 

\begin{figure}
   \centering
   \includegraphics[width=\columnwidth]{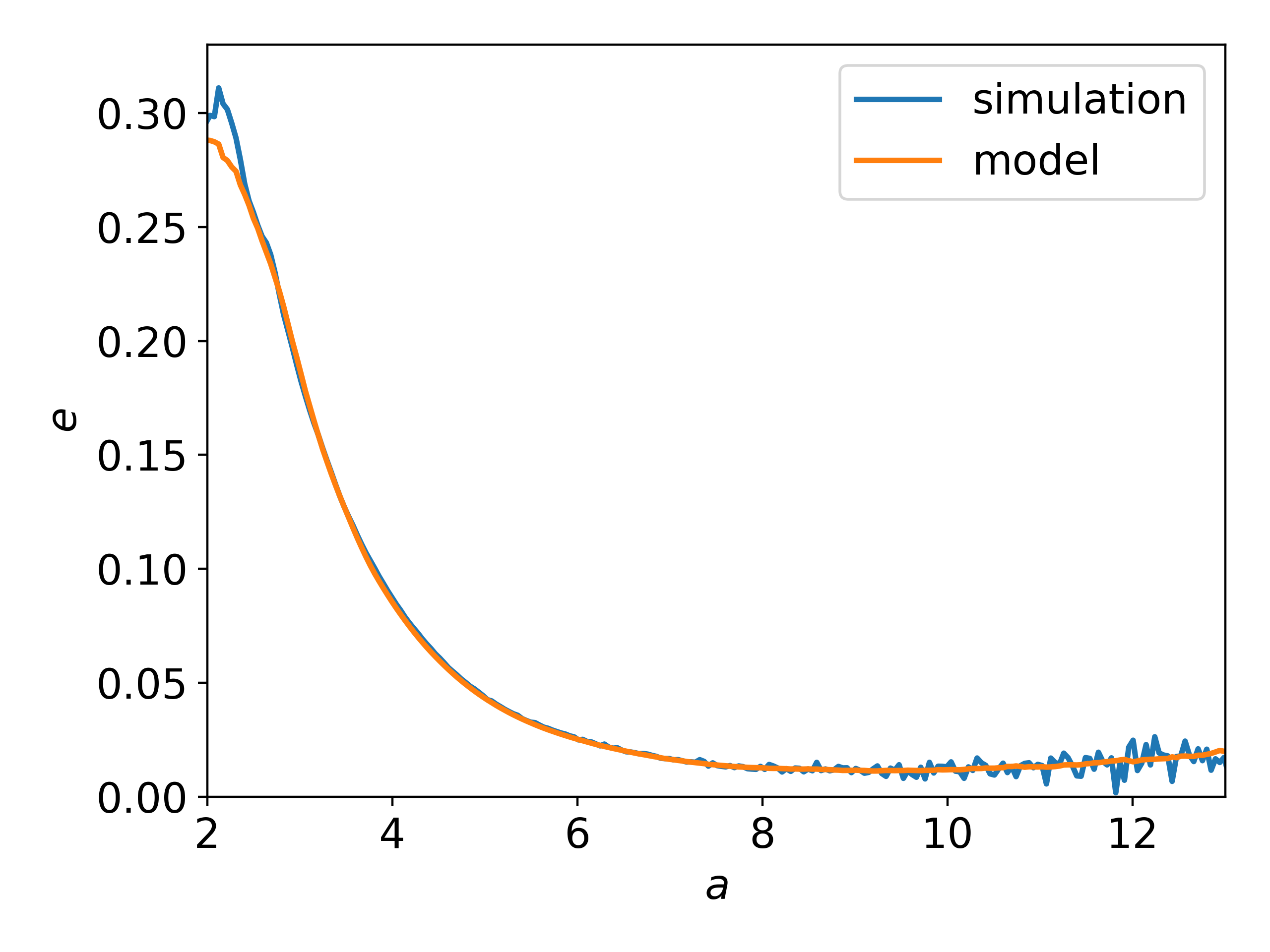}\\[-5pt]
   \includegraphics[width=\columnwidth]{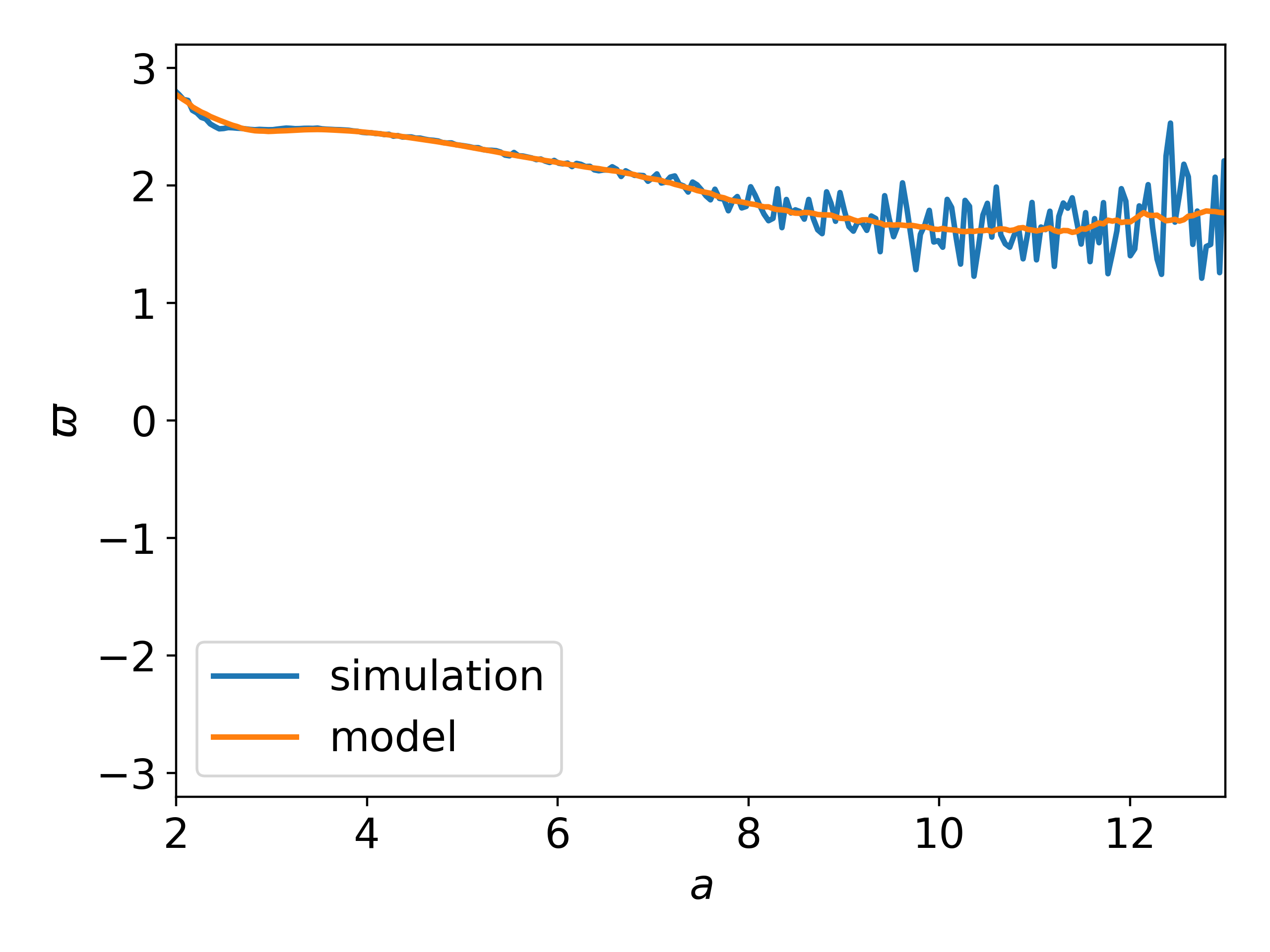}\\[-10pt]
   \includegraphics[width=\columnwidth]{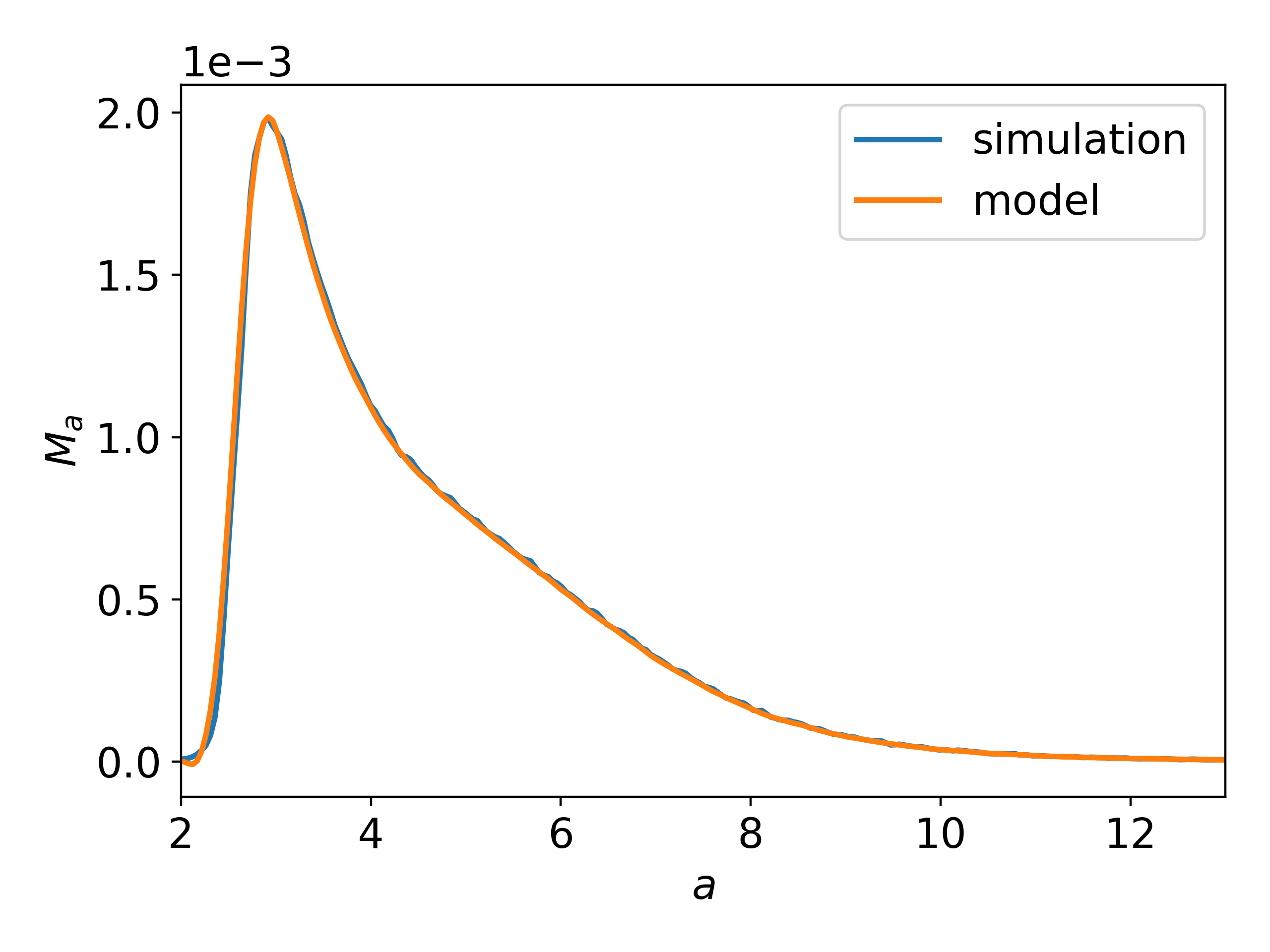}
   \caption{Profiles of $e(a)$ (top panel), $\varpi(a)$ (middle panel) and $M_a$ (bottom panel) from our reference simulation (blue curves) and corresponding filtered-interpolated profiles (orange curves). For clarity, we remark that the orange curves are obtained by filtering and interpolating with cubic splines the blue curve. The orange curve is used as an input to generate the theoretical model based on the numerical simulation.}
\label{fig:profiles}%
\end{figure}

In Fig. \ref{fig:profiles} we show the profiles extracted from our reference snapshot and the interpolated profiles we use for generating the analytical model. No appreciable differences can be found changing the parameter choice of the filter: as shown in Fig. \ref{fig:profiles}, the filtering process has the only effect of reducing the noise of the dataset but it does not alter appreciably its qualitative form.

We use Eq. (\ref{eq:vr}), (\ref{eq:vpress}) to calculate $v_{R,{\rm th}}$ and $v_{\theta{,\rm th}}$. We then solve Eq. (\ref{eq:veloverteq32}) to obtain the vertical scale height $H_{\rm th}$ and use it for calculating $v_{z,{\rm th}}$ from Eq. (\ref{eq:vz}). We finally use Eq. (\ref{eq:sigmamod}) to calculate the surface density morphology $\Sigma_{\rm th}$. Results are plotted as 2D maps where the region with $a<a_{\rm cav}$ has been excised -- where $a_{\rm cav}= 2.7 a_{\rm bin}$ is the semi-major axis where $M_a$ is maximum. This is done in order to exclude from our analysis the cavity area in the simulation, where the dynamics of the gas is strongly affected by the binary.

\begin{figure*}
   \centering
   \includegraphics[width=\columnwidth]{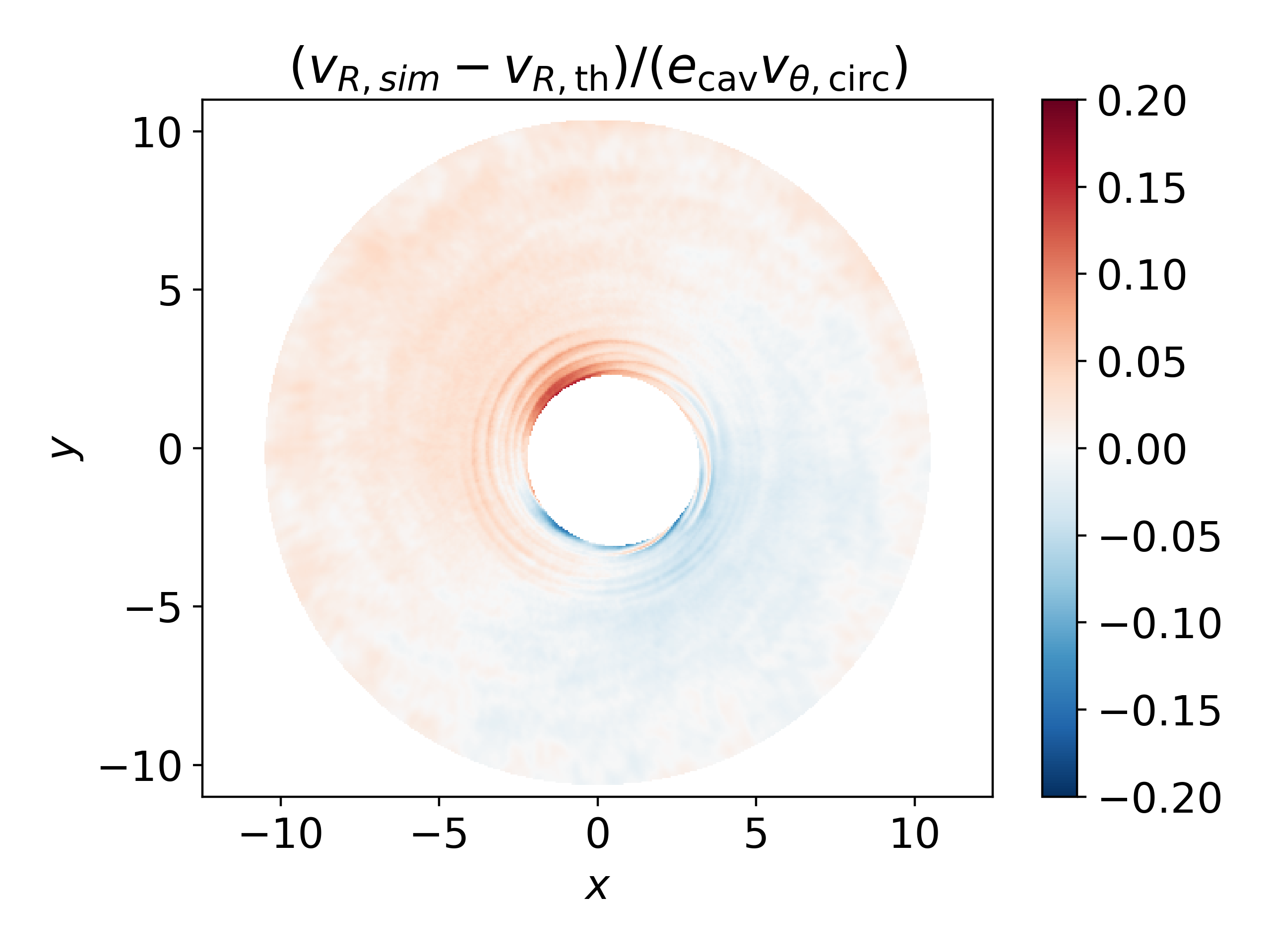}
   \includegraphics[width=\columnwidth]{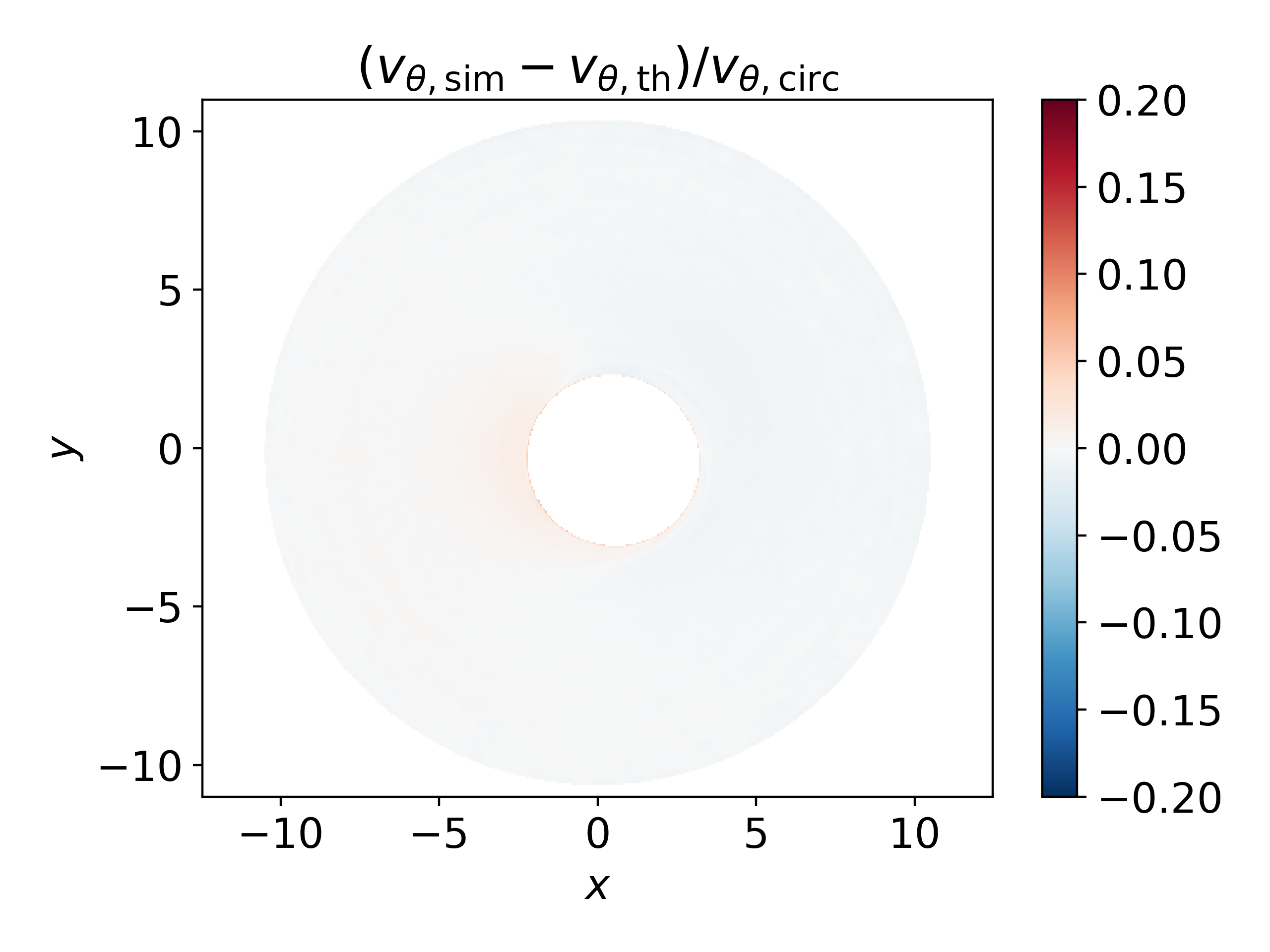}\\
   \includegraphics[width=\columnwidth]{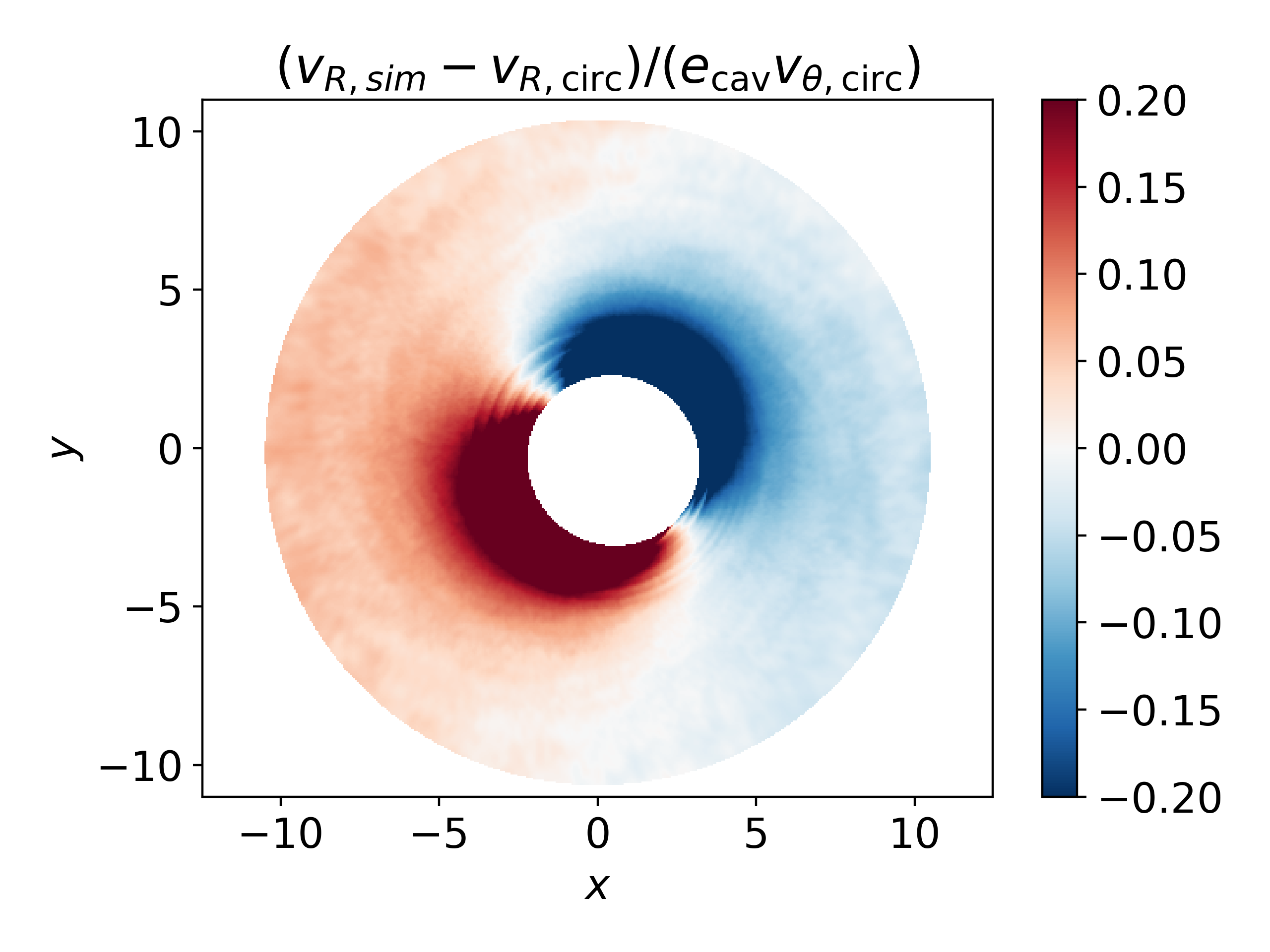}
   \includegraphics[width=\columnwidth]{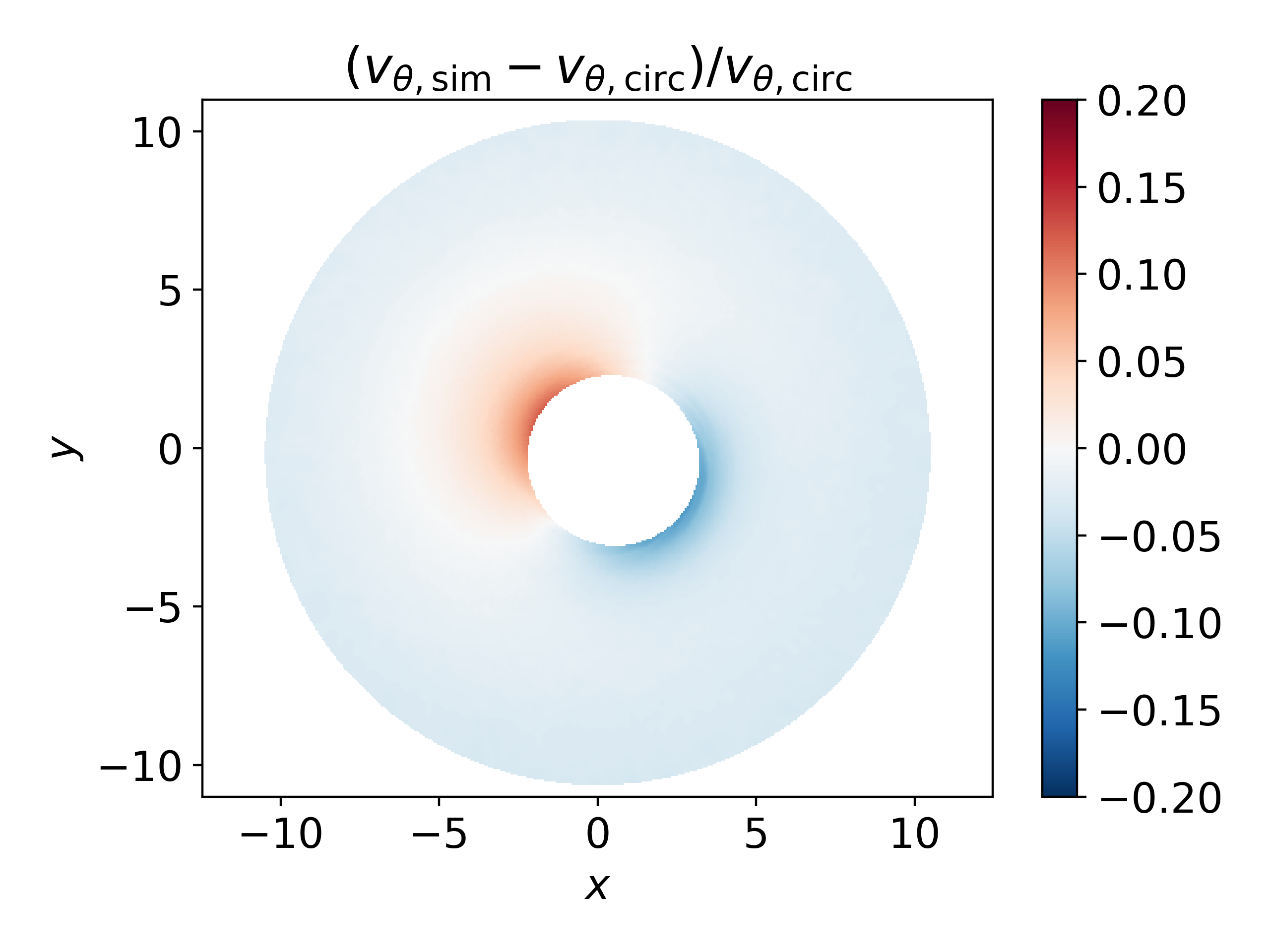}\\
   \caption{Top panels: comparison of the velocity fields $v_R$ and $v_\theta$ in the form of residuals between the simulation and the theoretical predictions in Eq. (\ref{eq:vr}) and (\ref{eq:vpress}); $(v_{R,{\rm sim}}-v_{R,{\rm th}})/v_{\theta,{\rm circ}}$ (top left panel) and  $(v_{\theta,{\rm sim}}-v_{\theta,{\rm th}})/v_{\theta,{\rm circ}}$  (top right panel), respectively. Bottom panels: similar to top panels but subtracting a circular Keplerian velocity profile ($v_{\theta,{\rm circ}}$) from the $v_{R,{\rm sim}}$ and $v_{\theta,{\rm sim}}$ in order to highlight the impact on the residuals of not accounting for the eccentric nature of the disc and of pressure corrections in the radial/azimuthal motion; $(v_{R,{\rm sim}}-v_{R,{\rm circ}})/v_{\theta,{\rm circ}}$ (bottom left panel), where obviously $v_{R,{\rm circ}}=0$, and  $(v_{\theta,{\rm sim}}-v_{\theta,{\rm circ}})/v_{\theta,{\rm circ}}$ (bottom right panel). In the bottom right panel it can be clearly seen that the simulation azimuthal velocity $v_{\theta,{\rm sim}}$ at large radii is sub-Keplerian, as $v_{\theta,{\rm circ}}$ does not account for the pressure support term.}
\label{fig:vrvphi}%
\end{figure*}

\begin{figure*}
   \centering
   \includegraphics[width=\columnwidth]{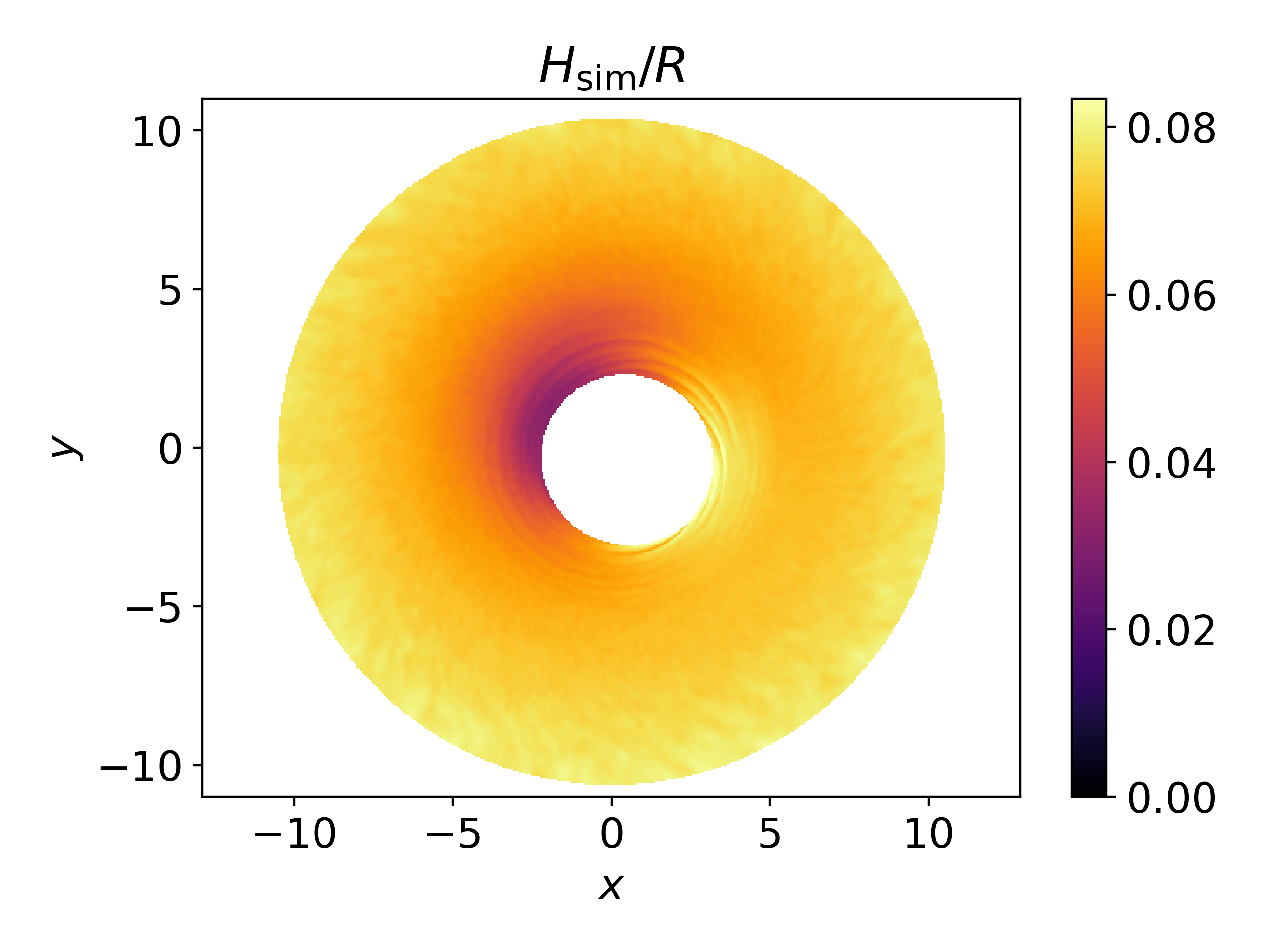}
   \includegraphics[width=\columnwidth]{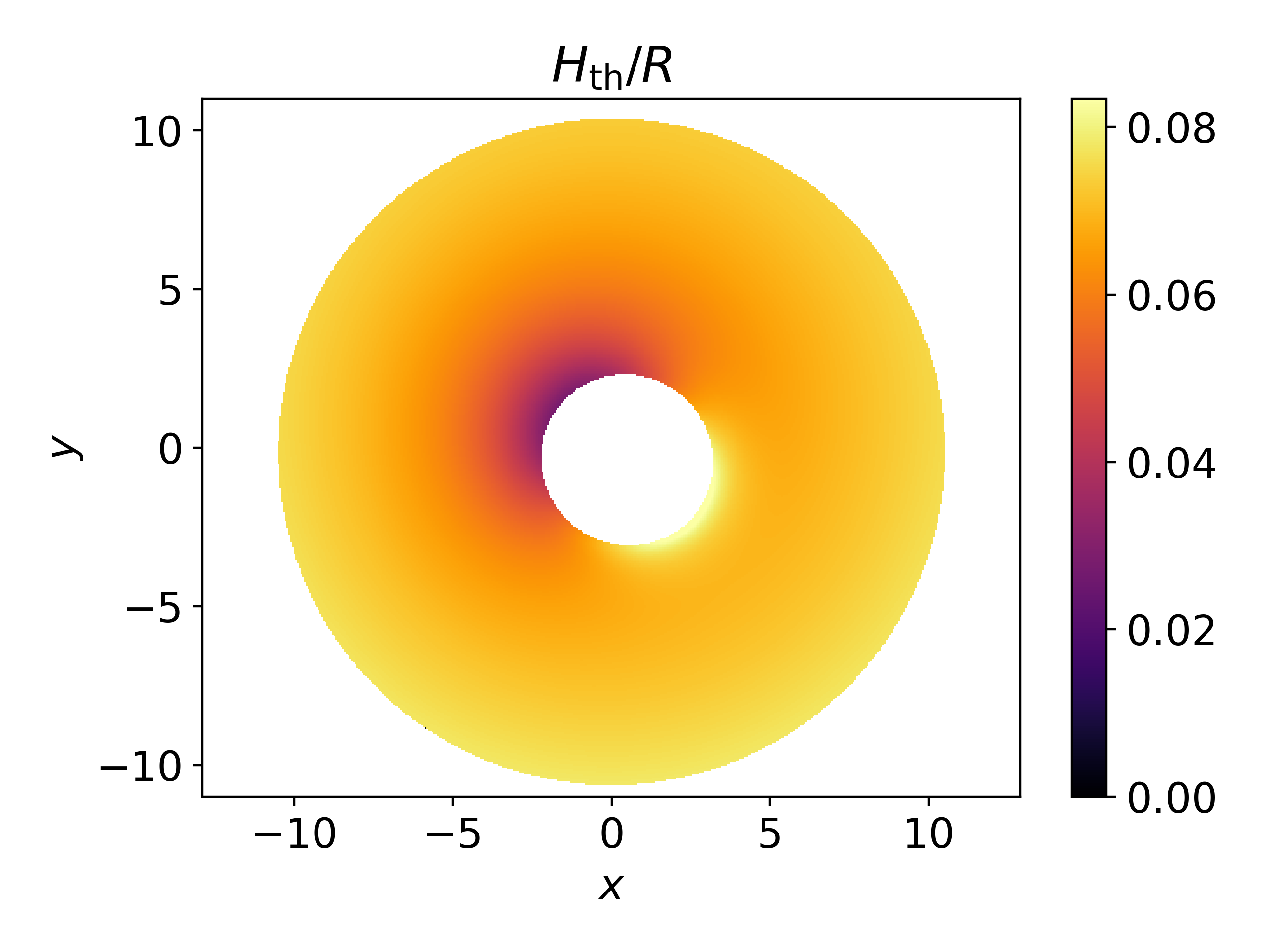}
   \caption{Left panel: Disc $H_{\rm sim}/R$ from simulation, calculated as $H_{\rm sim}/R=\langle z^2(R)\rangle^{1/2}/R$; note that $\langle z^2(R)\rangle=H_{\rm sim}$ by definition, being the vertical second moment of the density distribution. Right panel: disc $H_{\rm th}/R$ from solving Eq. (\ref{eq:veloverteq32}) using as input $e(a)$ and $\varpi(a)$ and $\cs(R)$ from the simulation.}
\label{fig:hor}%
\end{figure*}

\begin{figure*}
   \centering
   \includegraphics[width=\columnwidth]{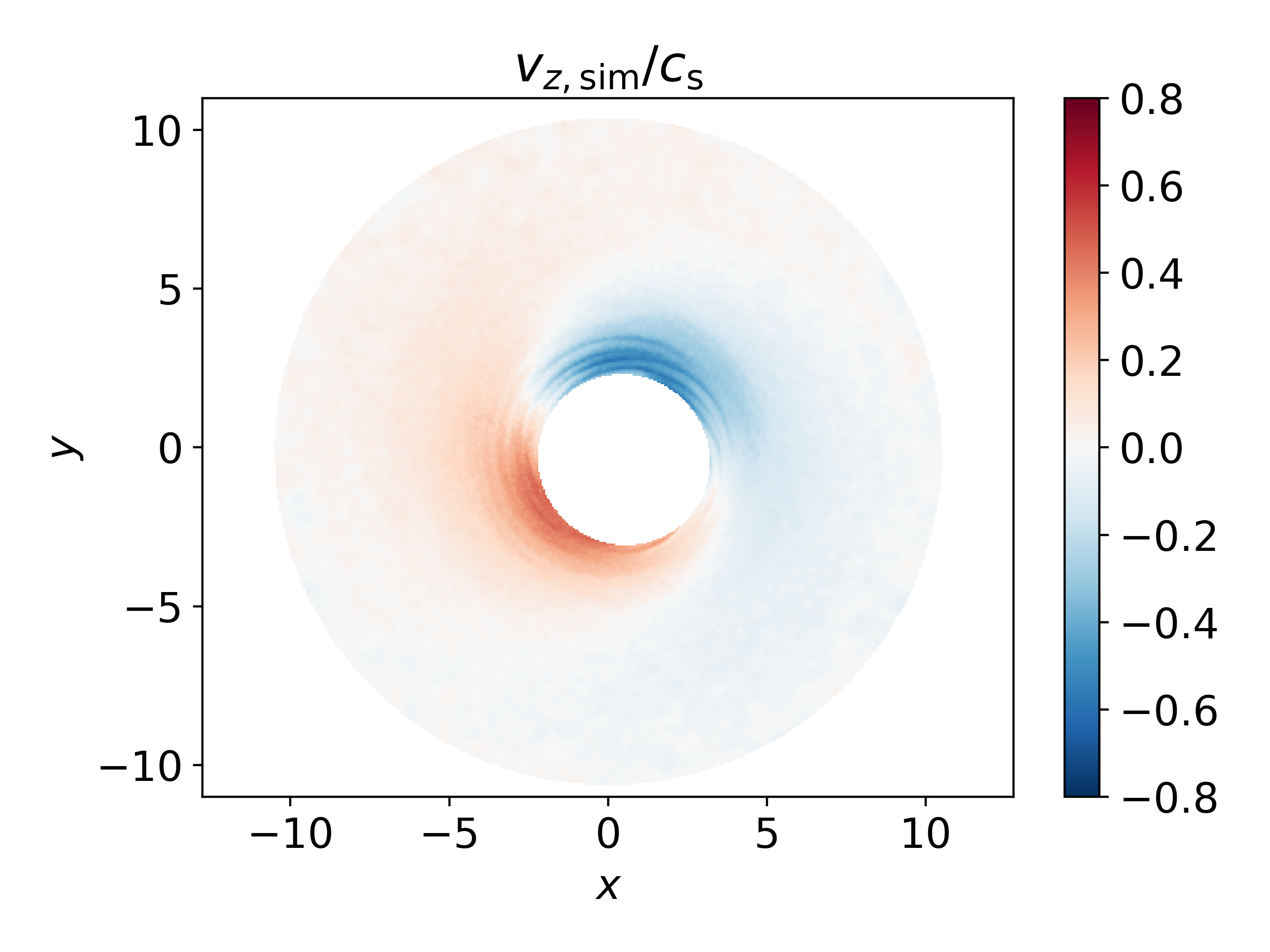}
   \includegraphics[width=\columnwidth]{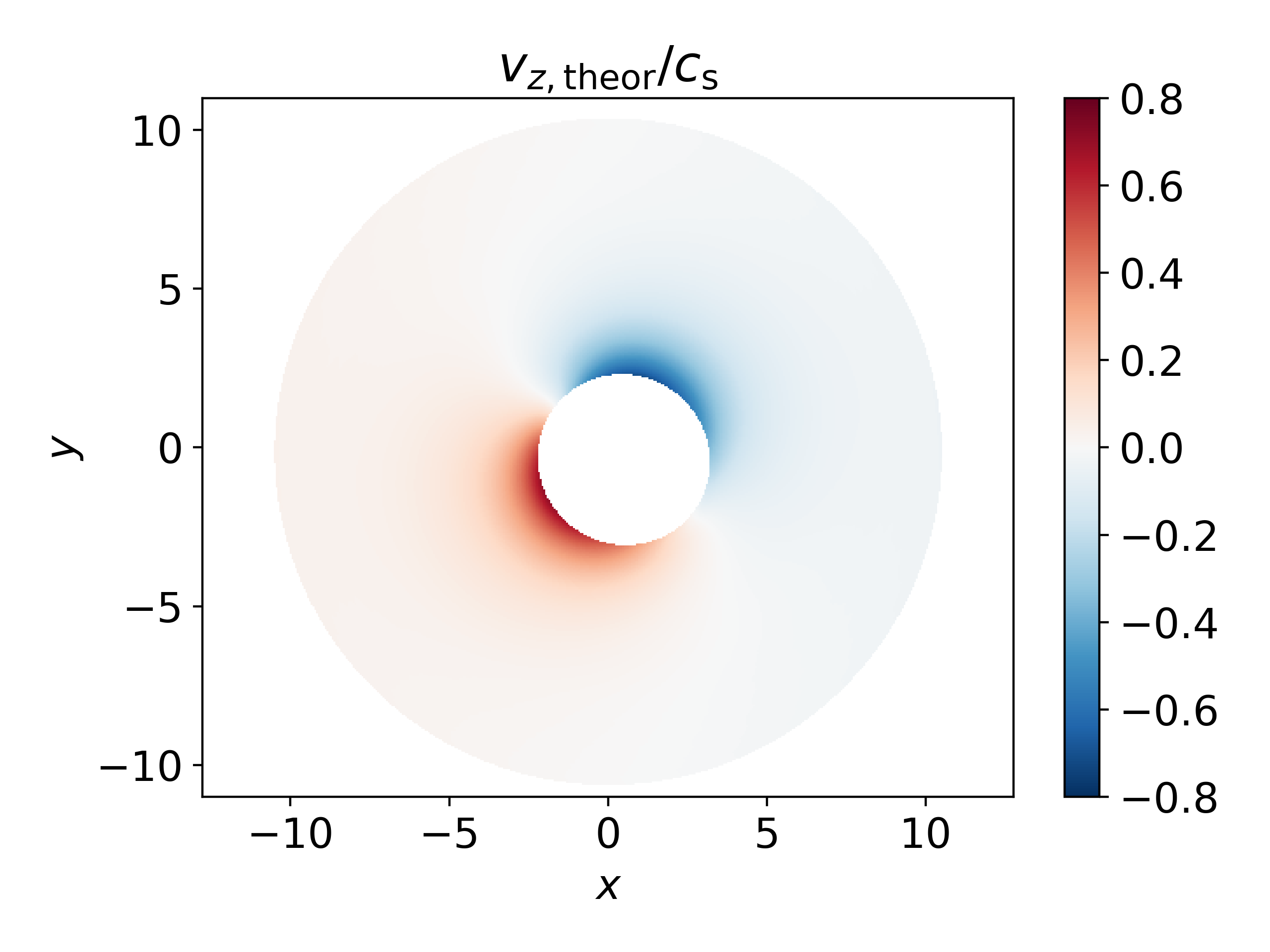}\\
   \includegraphics[width=\columnwidth]{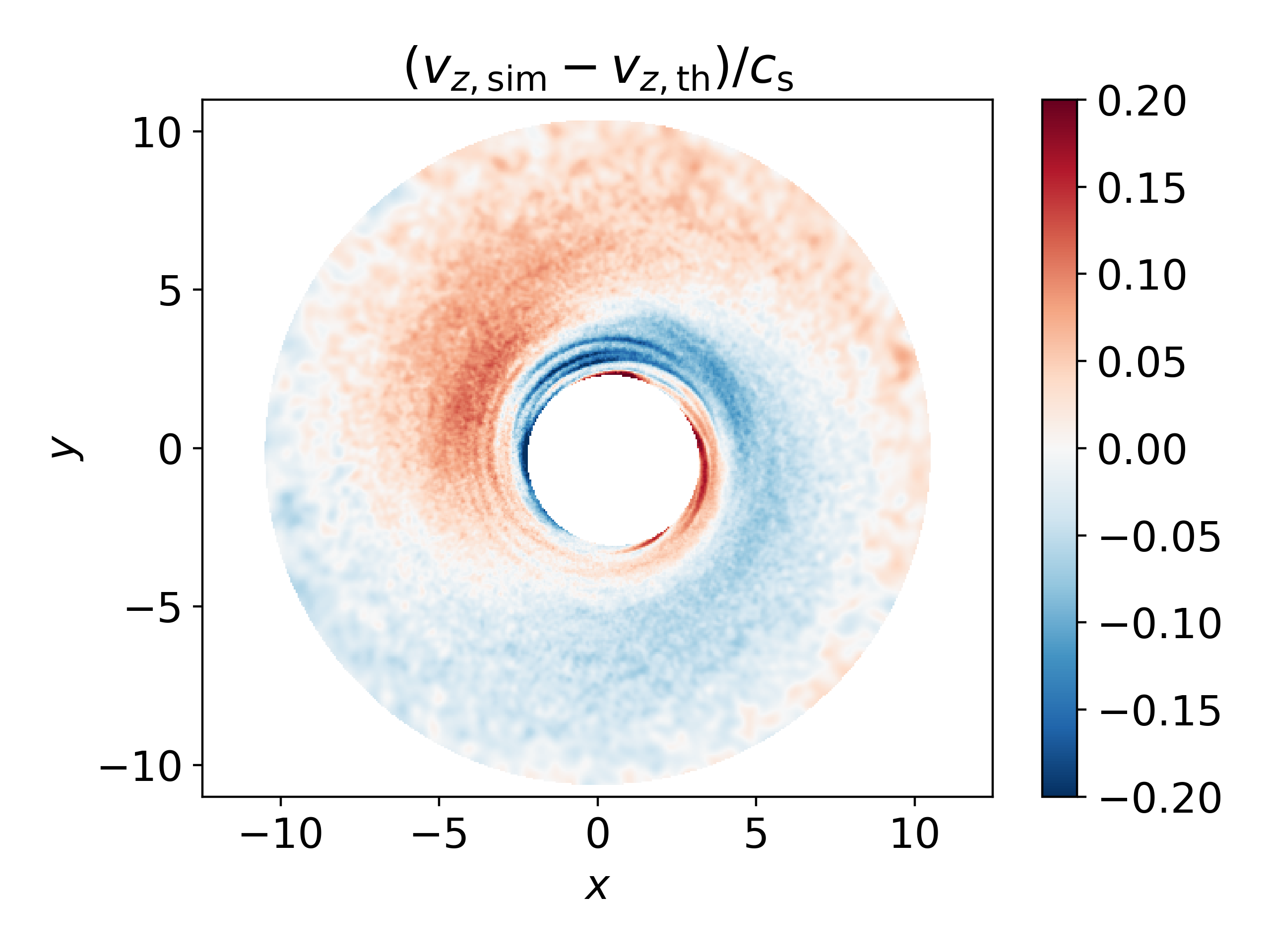}
   \includegraphics[width=\columnwidth]{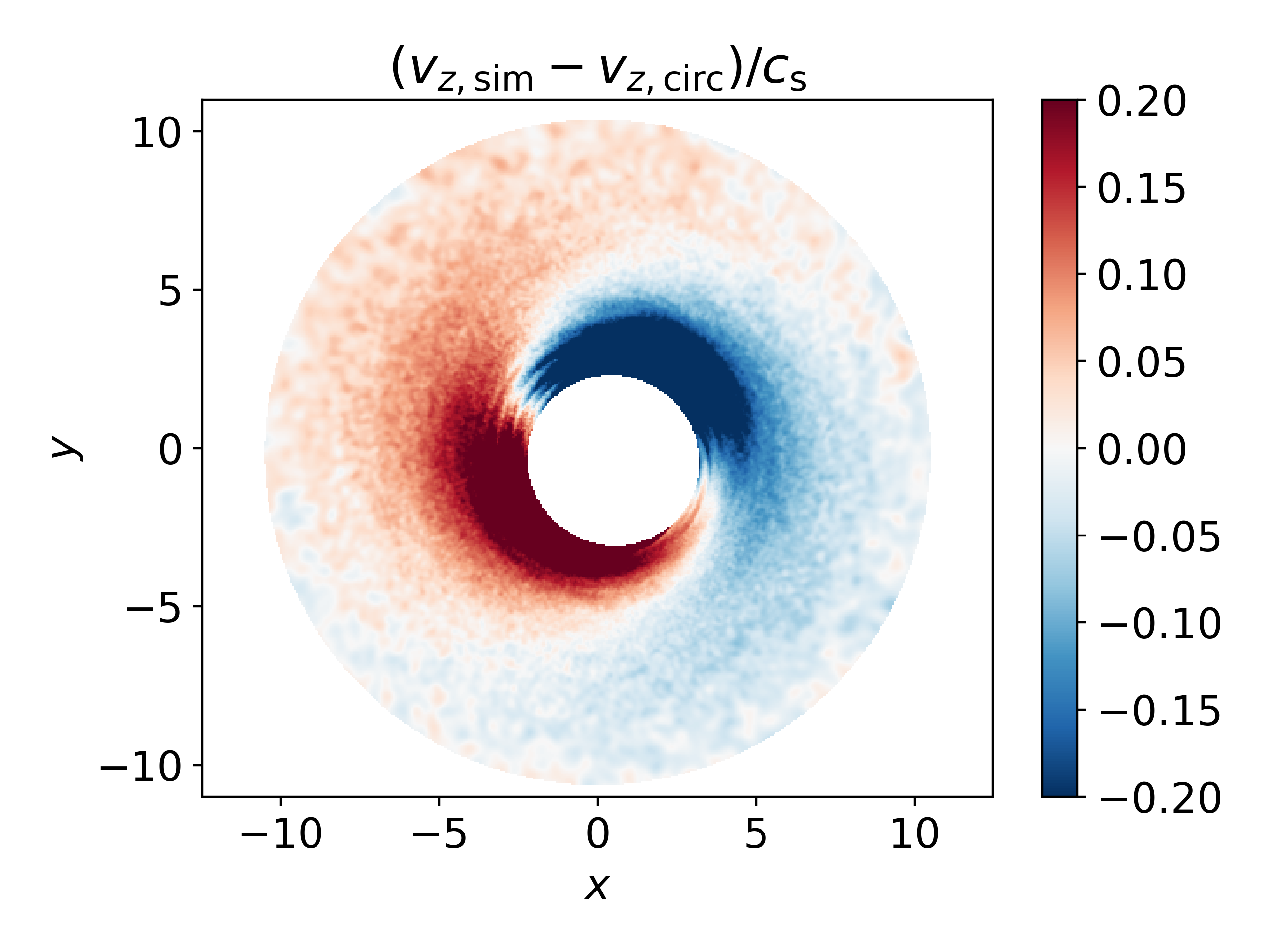}
   \caption{Top panels: $v_{z,{\rm sim}}/\cs$ velocity field from the simulation (top left panel) compared to the theoretical model $v_{z,{\rm th}}/\cs$ obtained solving Eq. (\ref{eq:veloverteq32}) (top right panel). Bottom panels: residuals of the simulation against the theoretical model $(v_{z,{\rm sim}}-v_{z,{\rm th}})/\cs$ (bottom left panel), i.e. the top left panel minus the top right panel of this figure, and with respect to a circular disc $(v_{z,{\rm sim}}-v_{z,{\rm circ}})/\cs$ (bottom right panel), i.e. $v_{z,{\rm circ}}=0$; this last plot is in fact the same as top left panel but with rescaled colorbar for a direct comparison with the other residuals plot in the bottom left panel. We note that in the simulation the $m=1$ spiral shaped vertical velocity feature standing out even after the subtraction of the eccentric theoretical model suggests that some additional physical processes are affecting the disc dynamics.}\label{fig:vz}%
\end{figure*}

\begin{figure*}
   \centering
   \includegraphics[width=\columnwidth]{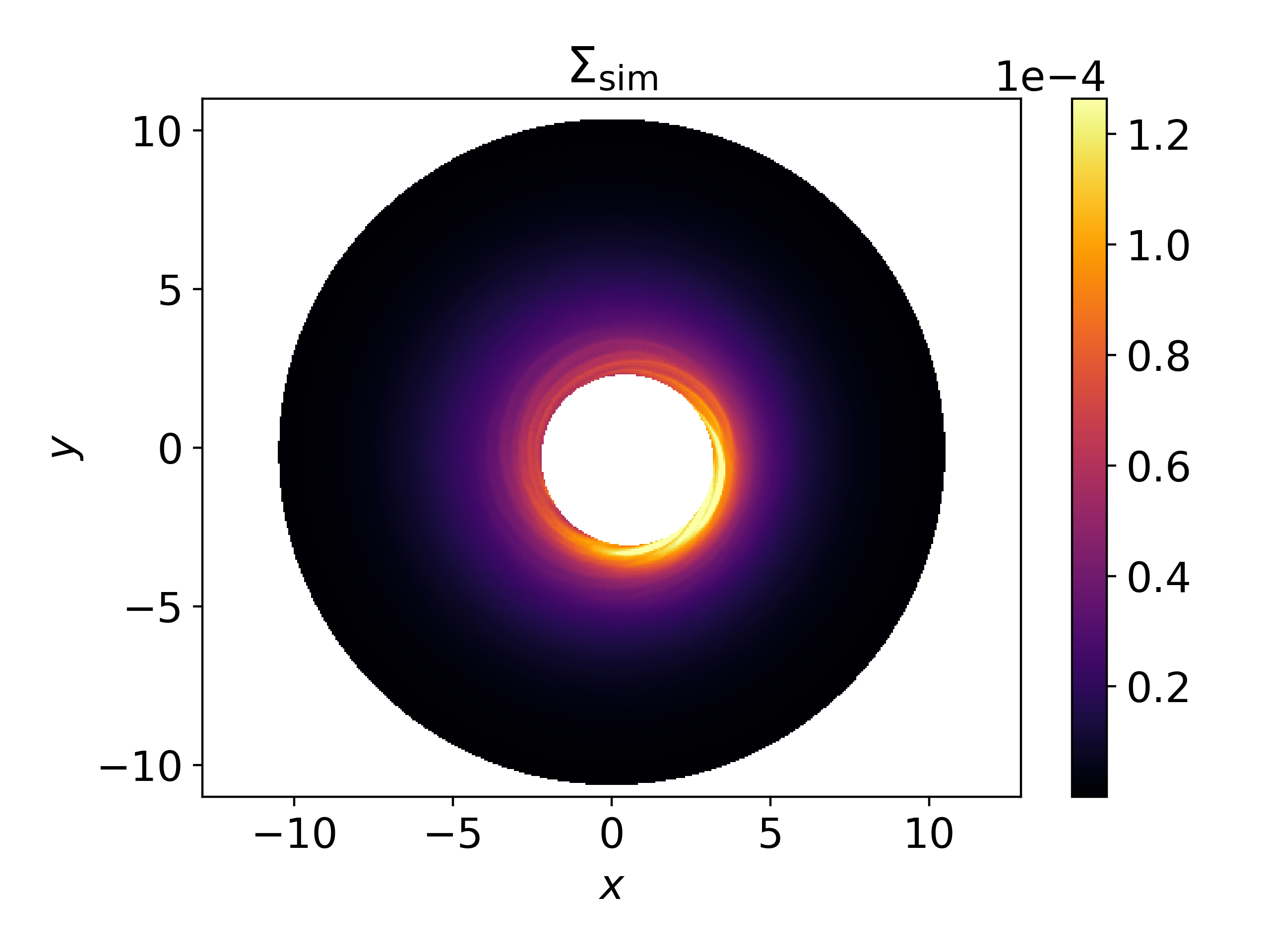}
   \includegraphics[width=\columnwidth]{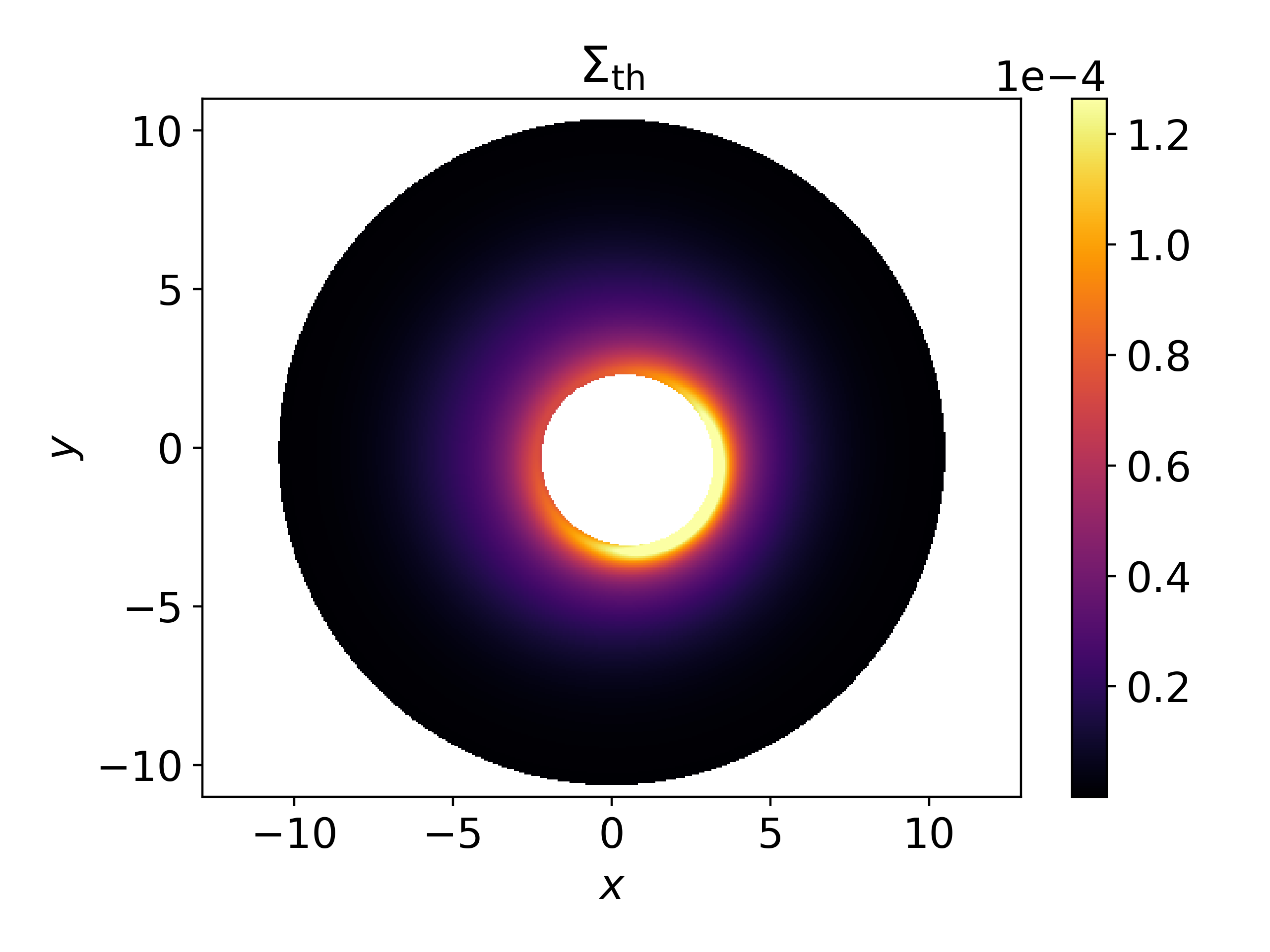}
   \caption{$\Sigma$ density field from the simulation (left panel) and relative analytical model (right panel) from Eq. (\ref{eq:sigmamod}).}
\label{fig:densityfield}%
\end{figure*}

\section{Numerical simulation vs the analytical model}\label{sec:discussion}

\subsection{Eccentric model vs circular}
We show a comparison between the simulation and the 3D analytical model discussed in the previous section in Fig. \ref{fig:vrvphi}, \ref{fig:hor}, \ref{fig:vz} and \ref{fig:densityfield}. Velocities $v_R$, $v_\theta$ and $v_z$ are presented in the form of residuals obtained subtracting the theoretical model from the corresponding maps computed from the simulation. For comparison we plot also the residuals obtained subtracting a circular Keplerian velocity field in order to highlight how large would be the systematic error of not accounting for the disc being eccentric, and the typical residual patterns obtained. We note that subtraction of kinematic models different from circular, non-tilted discs is not yet a common practice when analysing line emission observations.

More specifically, Fig. \ref{fig:vrvphi} shows a comparison between the normalised residuals  for $v_R$ and $v_\theta$. In the same figure, the residuals obtained subtracting $v_{R,{\rm circ}}=0$ and $v_{\theta,{\rm circ}}=\sqrt{GM_{\rm bin}/R}$. We normalise the residuals for $v_{\theta}$ with $v_{\theta,{\rm circ}}$, i.e. the circular Keplerian velocity at the $x-y$ pixel radius; we normalise $v_R$ using $e_{\rm cav}v_{\theta,{\rm circ}}$ where $e_{\rm cav}=0.2$, which represent the maximum $v_R$ achieved by the material at the cavity edge. 

Fig. \ref{fig:hor} shows a comparison between the disc aspect ratio $H/R$ in the simulation and the one calculated from Eq. (\ref{eq:veloverteq32}). 

Fig. \ref{fig:vz} shows a comparison between the vertical velocity field in the simulation $v_{z,{\rm sim}}$ against the one obtained from the theoretical model $v_{z,{\rm th}}$. In Fig. \ref{fig:vz} we also show the residuals from the subtraction of the two, and one where $v_{z,{\rm circ}}=0$ is subtracted. This allows us to perform a direct comparison of the order of magnitude of the residuals when vertical oscillations are neglected, e.g. when considering a circular disc model. We normalise the residuals for $v_z$ with $\cs$: as mentioned in footnote \ref{fn:csnorm}, that is a natural scale for eccentric vertical velocity perturbations at $z=H$.

Finally, in Fig. \ref{fig:densityfield}, we show a comparison between the surface density $\Sigma$ for both the simulation and the theoretical model, obtained from Eq. (\ref{eq:sigmamod}).

\subsection{Agreement between the simulation and the model}

We here discuss more quantitatively the agreement between the numerical simulation and the theoretical model results presented in the previous sections.

The residuals shown in Fig. \ref{fig:vrvphi} highlight that the azimuthal and radial motion in the simulation are well described by the theoretical model. 

Residuals of the $v_R$ field show a maximum residual value of $|\Delta v_R|_{\rm max}/(e_{\rm cav}v_{\theta,{\rm circ}})\sim 15\%$ located at the cavity pericentre, where a tidal stream of material is launched from the tidal interaction with the binary (which is not captured by the theoretical model). The average residual value is $|\Delta v_R|_{\rm avg}\lesssim 5\%$, the pattern of these residuals suggest that the disc wobbles around the centre of the numerical domain. This is expected, since the disc has finite mass, both the centre of mass of the disc and that of the binary orbit around the centre of mass of the binary + disc system (which by construction is located in the centre of the domain in $(x,y)=(0,0)$).
Not considering the corrections due to the disc eccentric nature for $v_R$, i.e. in a circular disc, produces $|\Delta v_{R,{\rm circ}}|_{\rm max}/(e_{\rm cav}v_{\theta,{\rm circ}})\sim 100\%$. The comparison with the circular case, shows as expected an increase of the radial velocity reaching its maximum in absolute value $|v_R|$ at true anomaly $f={\rm \pi}/2$ and $f=3{\rm \pi}/2$.

The $v_\theta$ field similarly has a maximum residual value of $\Delta v_\theta/v_{\theta,{\rm circ}}\sim 2\%$, again located at the cavity pericentre where the tidal interaction of the binary is strongest, resulting in a transfer of angular momentum and energy to the disc. In the comparison with the circular disc case, the residuals grow to $|\Delta v_{\theta,{\rm circ}}|/v_{\theta,{\rm circ}}\sim 15\%$. Also in this case, the comparison with the circular case, shows that the azimuthal velocity reaches its maximum at pericentre and minimum at apocentre, as expected.

Fig. \ref{fig:hor} shows a comparison between the disc vertical scale height in the simulation and the theoretical prediction in the form of the disc aspect ratio $H/R$. Both the numerical simulation and theoretical model highlight the precence of a ``breathing'' mode that causes the disc to compress towards the midplane at pericentre and an expansion at apocentre. The difference between the disc thickness at pericentre and apocentre at the edge of the cavity reaches $H_{\rm apo}/H_{\rm peri}\sim 3$, as visible in Fig. \ref{fig:Hvsphi} for eccentricity $e\approx e_{\rm cav}=0.2$.

The vertical velocity field $v_z$ associated with this expansion and compression at apocentre and pericentre, respectively, is plotted in Fig. \ref{fig:vz}. We can recognise a good overall agreement between the simulation and the theoretical model. When compared with the residuals for the circular case, where $v_{z,{\rm circ}}=0$, the theoretical eccentric disc model appears to perform much better in reproducing the numerical simulation, producing overall a reasonable agreement. 

However, some differences are evident and should not be ignored. In particular, an $m=1$ spiral feature in the numerical simulation appears not to be captured by the theoretical model. Furthermore, the $v_z$ field in the numerical simulation shows a small level of asymmetry with respect to the pericentre longitude: i.e. the positive $v_z$ lobe looks slightly different in shape with respect to the negative one. Finally, the simulation $v_z$ map appears to show a slight shift with respect to the pericentre and apocentre of the location where $v_z$ vanishes, in contrast with what predicted by the theoretical model. 

We identify two possible origins of these effects, that are not mutually exclusive: on the one hand, the binary nature of the central source of gravity in the simulation is responsible for the additional features; on the other hand, the additional features in the simulation might be captured by the eccentric disc theoretical model if we considered also the effect of viscosity when calculating the disc vertical dynamics (Eq. \ref{eq:vz} and \ref{eq:veloverteq32}). 

We explored this second possibility by including the bulk viscosity stress term in Eq. (\ref{eq:vertvel1}) -- the derivation of the final equation including bulk viscosity can be found in Appendix \ref{appendix:bulk}. A bulk viscosity component is expected to be spuriously present in any numerical scheme due to numerical dissipation.
As explained in the appendix, the effect of this additional viscous term produce some asymmetry in the vertical velocity field and a shift of the location where $v_z$ vanishes, consistently with the one observed in the simulation. However, the new theoretical prescription for $v_z$ fails in any case to reproduce the $m=1$ spiral pattern visible in the residuals of $v_z$. For this reason we decided to keep the model in a simpler form, and not to include the comparison with the viscous theoretical model for $v_z$, which is beyond the scope of the current paper.

In conclusion, we cannot identify the precise origin of the discrepancies in the predicted $v_z$ and the simulated one. Such a spiral feature in the $v_z$ field of the simulation appears to be related to the presence of the binary or to spurious dissipation resulting in an effective bulk viscosity. We postpone a thorough investigation about the origin of this discrepancy to a future work. 

Spiral features in the vertical $v_z$ have been previously described in planet-disc simulations by \citet{bae2021}, and have been referred to as ``buoyancy spirals''. Even though the spiral features in our simulations possibly share a similar dynamical ``trigger'', i.e. the presence of a secondary mass in the system, buoyancy effects cannot be excited in our vertically isothermal numerical simulations, implying that vertical thermal stratification is not a fundamental requirement for these features to arise, at least in the $q\gtrsim 0.1$ mass-ratio regime. To our knowledge, such spiral features in the vertical $v_z$ have not been documented before in the literature of locally isothermal circumbinary disc simulations.

More generally, a new analysis of the numerical set in \citet{ragusa2020} reveals that other simulations show similar spiral features in the disc vertical velocity field. We missed this at first because our standard approach to produce 2D maps of 3D simulations requires vertically averaging the fluid properties. Given the anti-symmetric nature of the perturbation with respect to the midplane, vertical integration of the $v_z$ 3D field produces $v_z=0$ maps in the entire disc $x\textrm{--}y$ domain. 

We remark that this perspective is particularly intriguing, in the light of the detection of a spiral kinematic feature possibly associated with vertical motion in the $^{12}$CO emission of the system HD142527 \citep{garg2022}, that features a well known circumbinary eccentric disc.

\section{Synthetic observations}\label{sec:MCRT}

In this section we investigate the physical magnitude of the eccentric kinematic features discussed in the previous sections trying to assess their observability. To do so we rescale the hydrodynamical simulation and perform Monte Carlo radiative transfer on it, in order to explicitly check whether the velocity perturbations are within the current observational capabilities.  

\subsection{Monte Carlo Radiative Transfer}\label{sec:MCRTMCRT}

We perform Monte Carlo Radiative Transfer simulations using the code \textsc{mcfost} \citep{pinte2006,pinte2009} to produce channel maps of the line emission from the 3--2 transition of the $^{12}$CO.
For this purpose, we rescale the numerical simulations for the binary to have a separation of $15$ au, by multiplying the spatial quantities by $x_{\rm sc}=15$. This results in a cavity size in the disc of $\approx 40$ au. The two stars are assumed to have masses $M_1=0.91\,M_\odot$ and $M_2=0.09\,M_\odot$ and to radiate as two black bodies with temperature $T_1=4200$ K and $T_2=2900$ K consistently with an age of $3$ Myr \citep{siess2000}, respectively. Following this assumption about the binary mass, we rescale the velocities by $v_{\rm sc}=29.7/\sqrt{15}$ to have the velocities in ${\rm kms}^{-1}$. The total gas mass is $5\, M_{\rm J}$, while the dust component is included reflecting the gas density distribution rescaled by a factor 100. We remark that our purpose is to produce mock synthetic observations of the eccentric kinematic features discussed in the previous sections. As a consequence, a self-consistent evolution of both gas and dust is beyond the scope of the paper. We assume a chemical abundance ratio\footnote{Our choice of $\rho_{^{12}{\rm CO}}/\rho_{\rm H}$ instead of the typical $\rho_{^{12}{\rm CO}}/\rho_{\rm H}=10^{-4}$ constitutes a conservative estimate for the vertical velocity. An optically thinner disc receives less contribution in the line emission from the highest disc layers, that have larger vertical velocities.} $\rho_{^{12}{\rm CO}}/\rho_{\rm H}=5\cdot 10^{-5}$. 

We here perform the radiative transfer analysis with the sole purpose of investigating the qualitative appearance of the gas kinematics, which is instead not affected by this assumption. 
A self-consistent treatment of the dynamics of the gas + dust mixture is beyond the scope of the paper. 

Dust opacities are calculated using Mie theory assuming astrosilicate grains \citep{weingartner2001}, with sizes ranging between $s=0.03\textrm{--}1000\, \mu m$ and a distribution $dn(s)/ds\propto s^{7/2}$.  

We use $N_\gamma=10^7$ photons for calculating the temperature structure, which is then used for determining the line emission properties; the final channel maps are then calculated through ray-tracing with $N_\gamma=10^7$ photons.

Angular scales and fluxes are calculated placing the source at a distance $d=130$ pc, consistently with the distance of the Taurus region. We study the appearance of the system for two inclinations: $i=0^\circ$ -- i.e. face-on -- and $i=30^\circ$. These values are chosen in order to study the detectability of the vertical oscillations and of the other eccentric features predicted in the previous sections for both face-on and for moderately inclined discs. 

The result of the radiative transfer is a datacube with ``infinite'' spatial resolution (set by the hydrodynamical simulation), while spectral resolution bins in the cube are $\Delta v=0.004\, {\rm kms}^{-1}$ for the face-on case and $\Delta v=0.01\, {\rm kms}^{-1}$ for $i=30^\circ$.

\subsection{Observability of the eccentric features}

With the purpose of extracting a projected velocity map of the disc surface from the radiative transfer data cube, we produce moment 1 maps (M1, i.e. channels are collapsed attributing to each pixel the brightness-weighted average of the velocity from brightness-velocity profile) and moment 9 (M9, i.e. maps where the channels are collapsed attributing the value of the ``brightest'' velocity in the brightness-velocity profile of each pixel) maps, shown in Fig. \ref{fig:zoomin0}. A selection of channel maps can be found for completeness in Fig. \ref{fig:channelmaps} in Appendix \ref{sec:channelmaps}. In order to qualitatively reproduce a real observation, we convolve the image spatially with a circular, Gaussian beam $\Delta s=0.05$ arcsec and the velocity with $\Delta v=0.05\,{\rm km s}^{-1}$ (i.e. within ALMA observational capabilities). 

It can be easily noted that the collapse of the channel maps into the moment 1 map significantly underestimates the vertical velocity component in the $i=0^\circ$ case compared to the moment 9. This can be understood by looking at the brightness-velocity plot in the  bottom panels of Fig. \ref{fig:zoomin0}. The emission in each spatial pixel of the data cube has a broad spectrum that depends on the projected velocity and temperature in the different disc layers encountered along the line of sight -- the temperature stratification typical of protostellar discs (i.e., hot on the disc surface, cold close to the disc midplane) explains also the secondary peaks that are visible in both $i=0^\circ$ and $i=30^\circ$ pixel spectra \citep{pinte2018b}. Producing an M1 map implies taking the brightness averaged velocity as the representative velocity of each pixel: with brightness profiles such as those shown in the right panels of Fig \ref{fig:zoomin0}, M1 maps strongly underestimate the velocity of the upper emission surface in the pixel.
M9 maps better capture the fact that the velocity in the upper layers is $v_z\neq 0$, but remain far from a precise estimate.

To strengthen this statement, in Fig. \ref{fig:highvel} we show the brightness map of the $0.72\, {\rm km s}^{-1}$ channel from the radiative transfer data cube of the $i=0^\circ$ simulation. We can clearly see that, even though the M9 map shows a maximum vertical velocity of $v_{z,{\rm max}}\approx 0.25\, {\rm kms}^{-1}$, higher velocity channels can be quite bright and detectable using ALMA. We also note that in the top layer, the two lobes have opposite velocities, while in Fig. \ref{fig:highvel} they appear almost equally bright with positive velocities: i.e. each pixel contains emission from both the top and bottom layers. See also the following section for a direct comparison of the moment maps with the upper emission surface kinematics.

Despite the poor performance of moment maps at portraying the upper emission surface kinematics, M9 maps show recognizable patterns of the disc's eccentric nature both for face-on discs and for inclined ones: in the form of vertical motion and asymmetry between the projected velocity at apocentre and pericentre, respectively.

In conclusion, both M9 and M1 maps are not representative of the velocities in the upper emission surface, so we cannot provide a direct comparison of the radiative transfer with the theoretical model. However, the typical eccentric features, i.e. double anti-symmetric vertical velocity pattern and asymmetry between apocentre and pericentre are recognisable. 
More sophisticated tools are required to properly reconstruct the geometry of the upper emission surface from channel maps (e.g.  \textsc{discminer}, \citealp{izquierdo2021,izquierdo2023}), in order to carry out a detailed comparison with the model.

In the light of these considerations, we will not attempt a direct comparison of the radiative transfer results with the theoretical model. Instead, we will limit our analysis to inclining the simulation data and compare them with an inclined theoretical model generated for this purpose in the next section.

\begin{figure*}
   \centering
   \includegraphics[width=\textwidth]{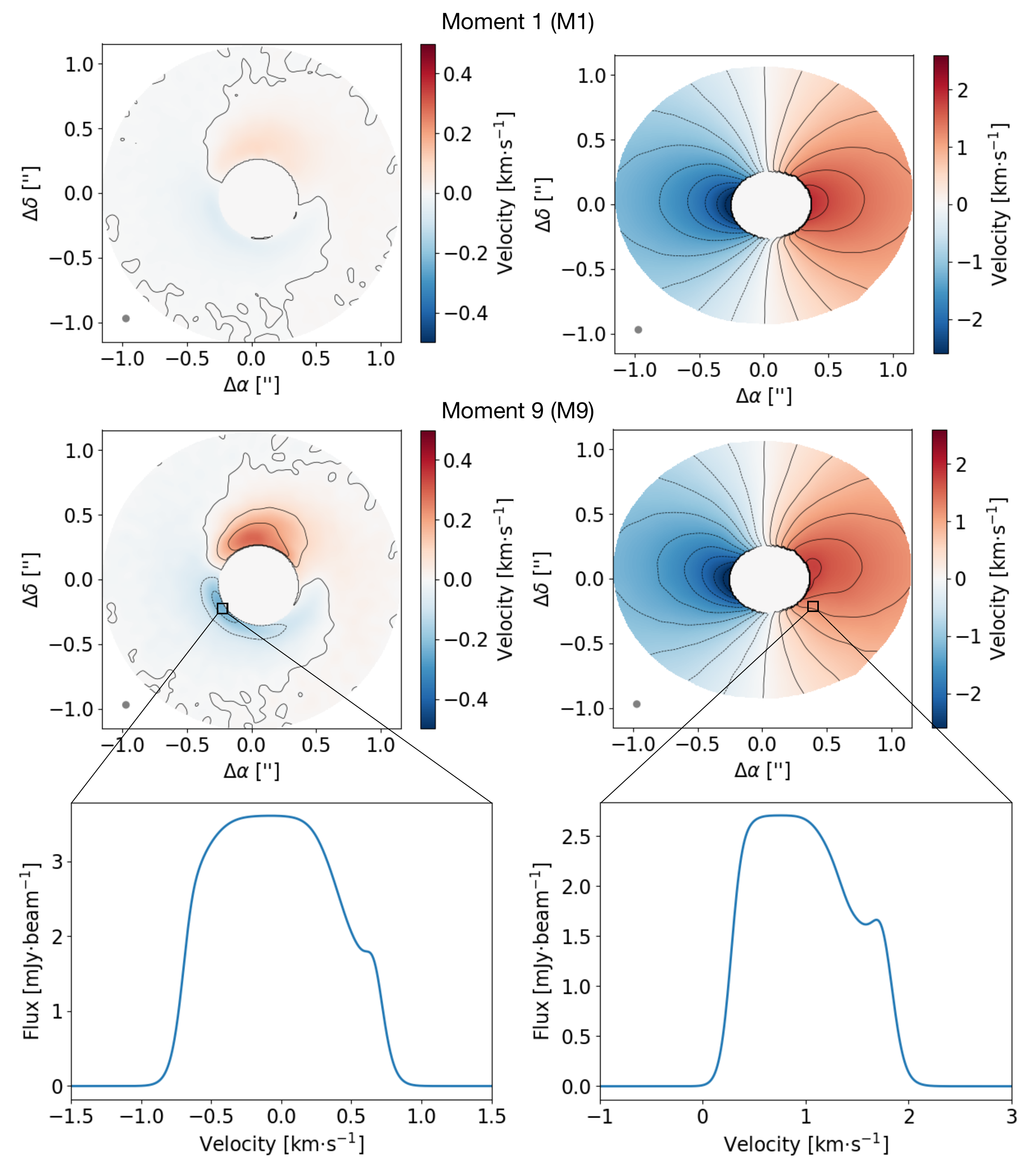}
   \caption{\textit{Top panels}: Moment 1 maps from Monte Carlo radiative transfer simulations $i=0^\circ$ (left) and $i=30^\circ$ (right). The grey circle in the bottom left corner represents the corresponding beam size.
   \textit{Middle panels}: Moment 9 maps, again obtained from the Monte Carlo radiative transfer of the simulation dump, $i=0^\circ$ (left) and $i=30^\circ$ (right). \textit{Bottom panels}: Brightness-velocity maps, $i=0^\circ$ (left) and $i=30^\circ$ (right), for selected pixels for highlighting the large span of different velocity contributions in individual pixels. The maps are obtained after convolving spatially the image with a Gaussian beam of $\Delta s=0.05$ arcsec (marked as a grey circle in the bottom left corner of each image) and $\Delta v=0.05\,{\rm kms}^{-1}$. Note that the convention about velocities in observations is that blue-shift of lines is associated with negative velocities, while red-shift is associated with positive velocities -- e.g., this implies that $v_z$ in this plot is opposite in sign with respect to that in Fig. \ref{fig:vz}. The case $i=0^\circ$ shows well how M1 maps tend to underestimate the vertical velocity contribution compared to M9 maps.}
\label{fig:zoomin0}%
\end{figure*}

\begin{figure}
   \centering
   \includegraphics[width=\columnwidth]{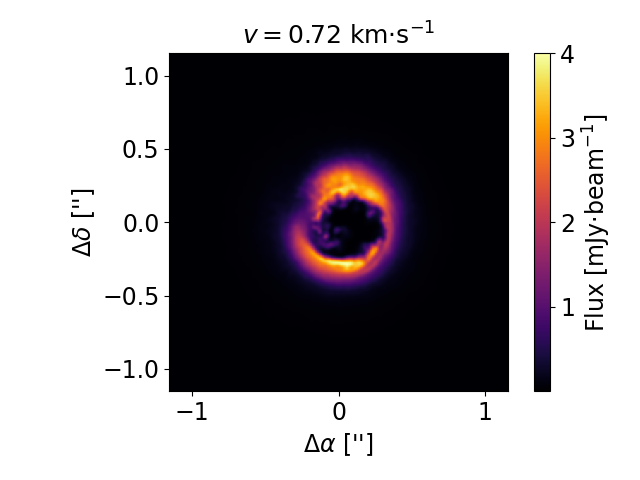}
   \caption{Brightness map of the $0.72\,{\rm kms}^{-1}$ channel from the data cube of the Monte Carlo radiative transfer, face-one $i=0^\circ$ inclination, performed on the simulation dump. The image clearly shows that, even though the M9 map in Fig. \ref{fig:zoomin0} shows maximum velocity of $\sim 0.25\,{\rm kms}^{-1}$ for vertical velocities, a bright contribution in higher velocities channels is also expected to be observable. }
\label{fig:highvel}%
\end{figure}

\subsection{Modelling the disc upper emission surface}\label{sec:tau=1}

In the previous section we saw we cannot directly compare the results from the radiative transfer moment maps with the theoretical model, as it is not possible to define precisely the velocities of the upper emission surface. Let us assume that we observe a circumbinary disc and that we have the tools to properly reconstruct from the channel maps the velocity projected on the observer's line of sight in the upper disc layers -- this is motivated by the current development of new observational tools for extracting the kinematics of the upper emission surface (\textsc{discminer}, \citealp{izquierdo2021}). How would the residuals from the subtraction of an eccentric disc model compare to those from a circular model? How strongly does the eccentric nature of the disc stand out compared to the circular model?

To answer these questions, we create a 3D model of the disc for both the theoretical models (eccentric and circular) and the numerical simulation, and we perform a direct comparison. The 3D model consists of spatial coordinates $\bm{\mathcal S}=\{x,y,z\}$ defining the 3D disc  upper emission surface, and velocities $\bm v=\{v_x,v_y,v_z\}$ as the velocity vector field of the material on it.

\begin{figure}
   \centering
   \includegraphics[width=\columnwidth]{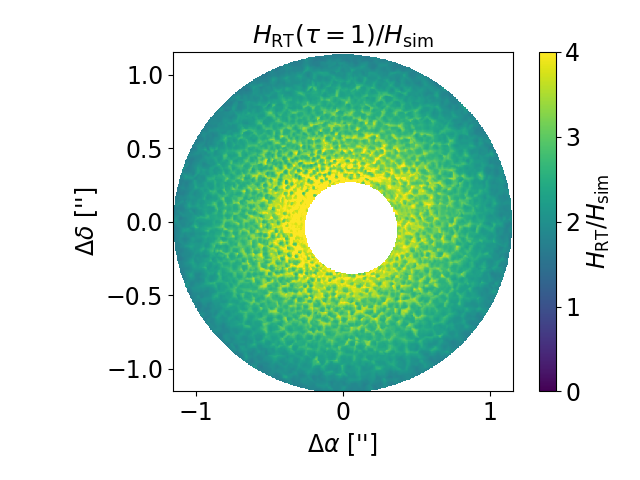}
   \caption{Map of the ratio between $H_{\tau=1}$, obtained from radiative transfer, tracing the upper emission surface and $H_{\rm sim}$, representing the simulation scale height. The ratio $H_{\tau=1}/H_{\rm sim}\approx 2\textrm{--}3$ across the disc. We consider $H_{\tau=1}\approx2.5H_{\rm sim}$ a conservative estimate of the location of the upper emission surface -- ``conservative'' since the higher in the disc atmosphere, the larger the vertical velocity.}
\label{fig:H_RT}%
\end{figure}

We first define the altitude of the disc upper surface using the altitude of the $\tau=1$ surface above the midplane $H_{\tau=1}$ output from the Monte Carlo radiative transfer simulation. The value of $\tau$ is calculated integrating along the line of sight the optical depth using the local maximum opacity $\max_\nu[k_\nu(x,y,z)]$  (i.e. the centre of the local line), where $k_\nu(x,y,z)$ is the local opacity, i.e. the surface $z(x,y)$ satisfying
\begin{equation}
\tau=\int_{+\infty}^{z(x,y)}\rho(x,y,z)\max _\nu[k_\nu(x,y,z)]{\rm dz}=1.\label{eq:tau}
\end{equation}
Such a surface qualitatively represents the highest elevation in the disc atmosphere at which the material contributes to the line emission, assuming the disc is face-on. We plot the altitude of the upper emission surface as $H_{\tau=1}/H_{\rm sim}$ for a comparison with the vertical scale height of the simulation in Fig. \ref{fig:H_RT}. The average value $\langle H_{\tau=1}/H_{\rm sim}\rangle\approx 2.5$, and it appears to be relatively constant throughout the entire disc. For this reason, we define our disc surface to be $z=2.5H_{\rm sim}$ for the simulation and $z=2.5H_{\rm th}$ for the circular and eccentric theoretical models.

Since Eq. (\ref{eq:tau}) is based on the maximum local opacity, it does not imply necessarily that the material at $z<H_{\tau=1}$ is not visible, as that would be the case only if $v_z={\rm const}$ throughout the entire gas column. Indeed, $\tau$ defined in Eq. (\ref{eq:tau}) should not be confused with $\tau_\nu$, which instead is a function of the frequency, and is obtained by integrating the opacity using the local $k_\nu$ (and not its local maximum as done for $\tau$). Since the opacity $k_{\nu}$ of the material changes with the material velocity, $\tau_\nu=1$ qualitatively defines a set of iso-velocity surfaces (one for each value of $\nu$, i.e. one for each velocity channel), probing different altitudes above the mid-plane. In fact, different layers of the disc atmosphere along the line of sight contribute to the line emission at different frequencies, despite lying at altitudes $z<H_{\tau=1}$.  

The disc has a top and a bottom surface, thus for our 3D eccentric and circular model we define them to be $\bm{\mathcal{S}}_{\rm top}=\{x,y,2.5H_{\rm th}\}$ for the top surface and $\bm{\mathcal{S}}_{\rm bottom}=\{x,y,-2.5H_{\rm th}\}$. 

We define the velocity field on the disc surfaces as $\bm v_{\rm top}=\{v_x,v_y,v_z\}$ for the top layer and $\bm v_{\rm bottom}=\{v_x,v_y,-v_z\}$ for the bottom layer, where $v_x$ and $v_y$ are obtained projecting $v_R$ and $v_\theta$ from Eq. (\ref{eq:vr}) and (\ref{eq:vpress}) on to the $x$ and $y$ unit vectors, while $v_z$ is calculated from Eq. (\ref{eq:vz}), calculated at $z=2.5 H_{\rm th}$. We produce a model with the same characteristics also for a circular disc, by projecting $v_{\rm circ}$ on to Cartesian unit vectors.

Similarly, after rescaling the numerical simulation as described in Sec. \ref{sec:MCRTMCRT} ($a_{\rm bin}=15$ au, $M_{\rm bin}=1\,{\rm M_\odot}$, $d=130$ pc, i.e. $x_{\rm sc}=15$, $v_{\rm sc}=29.7/\sqrt{15}$), we define the upper surface in the simulation to be $\bm{\mathcal{S}}_{\rm sim}=x_{\rm sc}\{x,y,2.5H_{\rm sim}\}$ and $\bm v=v_{\rm sc}\{v_{x,{\rm sim}},v_{y,{\rm sim}},2.5v_{z,{\rm sim}}\}$, where, we note, we applied an additional scaling of a factor 2.5 to the original $z_{\rm sim}$ and $v_{z,{\rm sim}}$, to account for the higher elevation of the upper emission surface compared to the vertical scale-height set by pressure -- this operation exploits the fact that $v_z\propto z/H_{\rm th}$. 

\subsection{Morphology of projected velocity residuals in circumbinary discs}

Consider an observer located at $z=+\infty$, and define the projected velocity $v_{\rm proj} =-v_z$ (following the convention that redshifted velocities are positive). At this point, the observer, looking down on the $x-y$ plane observing $v_{\rm proj}$ will see the kinematics and geometry of a face-on disc. We do this for both the eccentric and circular models, which constitute our face-on models. For the eccentric model the line of apses inherits its orientation from the simulation.

Inclined models are generated by rotating the face-on discs (described by $\bm{\mathcal{S}}$ and $\bm{v}$), keeping the observer (and associated coordinate system) fixed. First we perform a rotation around the $z$-axis, which sets the longitude of pericentre (this step is omitted in this work, meaning the inclined discs inherit the longitude of pericentre of the face-on disc). We then rotate around the $x$-axis by $i=30^{\circ}$ , resulting in an inclined disc with its line of nodes aligned with the $x$-axis. The longitude of pericentre is unchanged by this rotation, however the projected line of apses will be shifted towards the line of nodes as a result of foreshortening. Finally, without loss of generality, we set the position angle, $\mathrm{PA}=-90^{\circ}$ (calculated from the North anti-clockwise), as the position angle has no effect on the imaged disc.

In general, the kinematics of inclined eccentric disc is affected by two angles: the inclination, which sets the relative contribution of the vertical and horizontal velocity; and the argument of periapsis, which sets the relative contribution of the on-apse and off-apse velocities to the horizontal velocities. For symmetric beams, considered here, the position angle has no effect on the image.

We now have three different models of the geometry and projected velocity kinematics of the upper emission surface of a disc inclined by $i=30^\circ$. One for the circumbinary disc simulation, the other two for the eccentric and circular disc theoretical models.

\begin{figure*}
   \centering
    \includegraphics[width=\columnwidth]{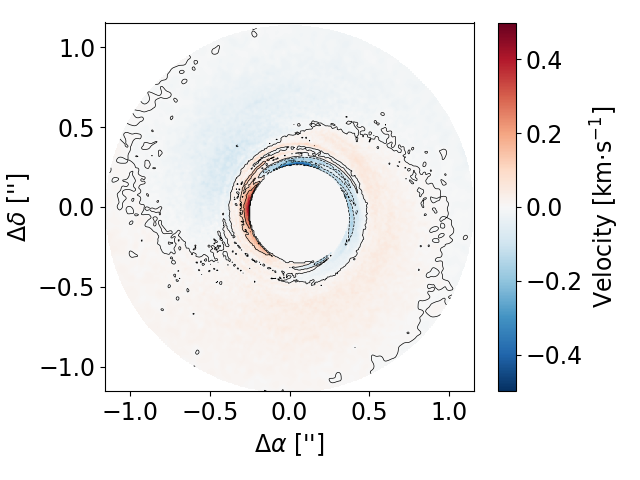}
   \includegraphics[width=\columnwidth]{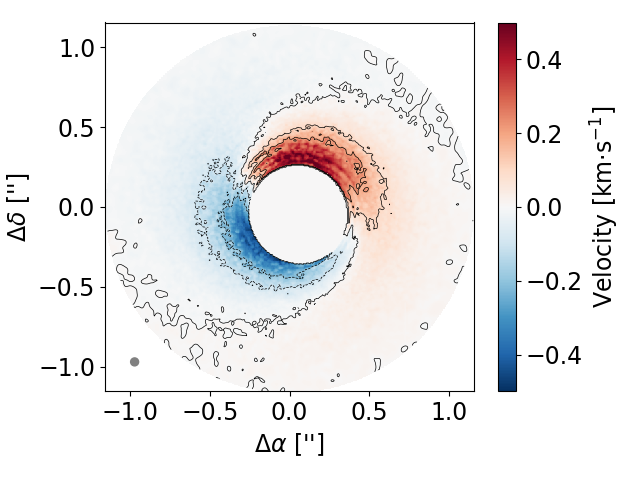}\\
   \includegraphics[width=\columnwidth]{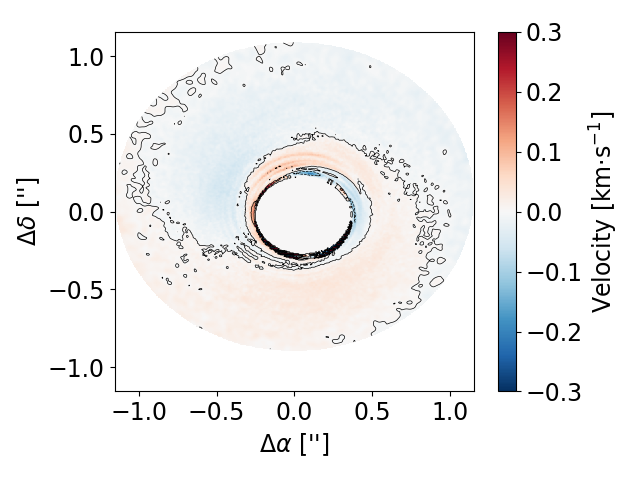}
   \includegraphics[width=\columnwidth]{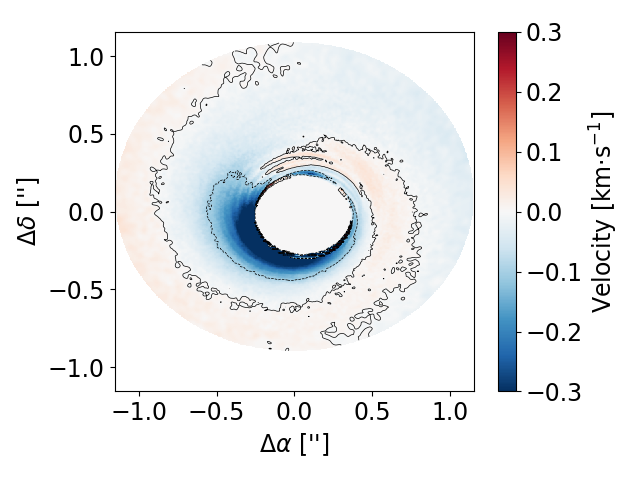}
   \caption{Residuals of the subtraction of the the eccentric theoretical model (left panels) and circular model (right panels) from the projected velocity in the simulation for $i=0^\circ$ (top panels) and $i=30^\circ$ (bottom panels). These maps were obtained assuming the upper emission surface to be $H_{\tau=1}=2.5H$ (as discussed in Sec. \ref{sec:tau=1}), and rescaling the simulations to have a binary with $a_{\rm bin}=15$ au, $M_{\rm bin}=1\,{\rm M_\odot}$, placed at a distance of $d=130$ pc. Similarly to Fig. \ref{fig:zoomin0}, we note that the convention about velocities in observations is that blue-shift of lines is associated with negative velocities, while red-shift is associated with positive velocities -- e.g., this implies that $v_z$ in this plot is opposite in sign with respect to that in Fig. \ref{fig:vz}. We note that the top right panel residual is obtained by subtracting $v_{z,{\rm circ}}=0$ (as the circular case does not have vertical motion) from $v_z$, thus this residuals map portrays the vertical velocity map on the $\tau=1$ surface for the case $i=0^\circ$. This highlights that the amplitude of the vertical oscillations reaches $v_z\sim 0.4\,{\rm kms}^{-1}$ in the upper emission layer.}
\label{fig:residuals}%
\end{figure*}

For the analysis described in this section, we consider only $\bm{\mathcal{S}}_{\rm top}$, under  the assumption we made above that we can isolate the kinematics coming exclusively from the top emission layer. We subtract both the ``eccentric theoretical model'' and the ``circular theoretical model'' from the simulation, the residuals of these two operations are shown in Fig. \ref{fig:residuals}.

A large, trailing, $m=1$ spiral with a velocity residual of $0.15\,{\rm km s}^{-1}$ appears in the $v_{\rm proj}$ kinematics (i.e. detectable with ALMA) when subtracting the eccentric theoretical model and the circular model. The eccentric theoretical model however appears to perform well in capturing all the other eccentric kinematic features that are not reproduced by a purely circular model. Indeed, residuals from the subtraction of the circular model from the simulation show instead a broad single-lobed azimuthal kinematic feature at the edge of the disc cavity with $\Delta v>0.3 {\rm kms}^{-1}$. Interestingly, such a feature does not have the appearance one might naively expect when subtracting a circular model from an eccentric disc: a double lobed feature (two lumps in the residuals with opposite signs) due to the faster and slower velocity of the material at pericentre and apocentre, respectively.
However, the structure of the residuals will change when orienting the system in a way that results in the disc having a different argument of periapsis (e.g., it is possible to get a two lobed feature with a different disc orientation).

In the light of the results presented here, it is important to consider the possible eccentric nature of the disc for explaining single and double-lobed structures in the projected velocity residuals. We also remark that we cannot be conclusive about the origin of the spiral-shaped velocity feature, not captured by the eccentric disc model. However, when observed, it might hint at the presence of a hidden (sub)stellar companion (if undetected).

\section{Conclusion}\label{sec:conclusion}

In this paper we discussed the morphology and kinematics of eccentric discs, with the final goal of providing observational signatures, aiding their identification in observational campaigns. To do so, we presented a theoretical model that takes as input information only three functions: namely, the disc eccentricity profile $e(a)$, the longitude of pericentre profile $\varpi(a)$, the derivative of the disc mass distribution $M_a(a)$, and an assumption about the locally isothermal disc sound speed $\langle \cs\rangle (a)$. Eccentric discs show a number of characteristic features:

(i) The azimuthal and radial velocities $v_\theta$ $v_R$, along with the surface density $\Sigma$ vary around the orbit  (Eq. \ref{eq:vr}, \ref{eq:vpress} and \ref{eq:sigmamod}). 

(ii) The changing vertical gravity around the eccentric orbit excites a vertical oscillation, corresponding to a variation of the disc scale height $H$ around the orbit (Eq. \ref{eq:veloverteq32}), and results in a vertical velocity $v_z \propto z/H$ (Eq. \ref{eq:vz}).

iii) The theoretical model explains many features of a numerical simulation of an eccentric circumbinary disc. However, a residual, $m=1$, spiral pattern in $v_z$ is not captured by the theoretical model. We speculate this could be produced by the binary potential or the effect of viscosity, neither included in our analytical treatment. The origin of this feature will be explored in future studies. We estimate this residual velocities in the spirals to be of the order of $\Delta v_z\approx 0.15\,{\rm kms}^{-1}$.

iv) We perform radiative transfer Monte Carlo on our numerical simulation snapshot to obtain information about $^{12}{\rm CO}$ line emission for inclination $i=0^\circ$ (face-on) and $i=30^\circ$. We find that moment 1 maps are not suitable to highlight the vertical motion in the $i=0^\circ$ case, since $v_z$ is anti-symmetric with respect to the mid-plane. Better results can be obtained with moment 9 maps.

v) Both $i=0^\circ$ and $i=30^\circ$ radiative transfer M9 maps show features of eccentric disc evolution that are within ALMA observational capabilities: the first showing a vertical velocity contribution that is is not present in circular face-on discs; the second showing an asymmetric butterfly diagram of the iso-velocity contours, indicating different velocity at pericentre and apocentre. 

vi) In the face-on case ($i=0^\circ$), the M9 map reveals the vertical velocity reaches a maximum of $v_{z,{\rm max}}\sim 0.25\,{\rm kms}^{-1}$. However, channel maps show significant signal at higher velocities from the upper layers of the disc, making $v_z$ eccentric signatures potentially detectable in face-on discs even for eccentricities $e\ll0.2$ (e.g., $v_{z,{\rm max}}\sim 0.4\,{\rm kms}^{-1}$ on the $\tau=1$ surface in Fig. \ref{fig:residuals}). 

vii) The subtraction of a circular Keplerian disc model from an eccentric disc's kinematics can, counterintuitively, produce a single lobed pattern in the projected velocity map. Eccentric disc kinematics should be considered when patterns such as those visible in Fig. \ref{fig:residuals} are observed. However, other patterns are possible for different disc orientations.

In future works we will further investigate the origin of the spiral velocity pattern discussed in (iii), determining whether such a feature can be considered a smoking gun signature of the presence of a binary (sub)stellar companion or a feature of any eccentric disc. In any case, the results presented in this paper represent an intriguing starting point for the interpretation of similar spiral patterns in the vertical velocity field that are starting to be observed (e.g. \citealp{garg2022}).

\begin{acknowledgements}
We thank the anonymous referee for their comments to the manuscript.
E.R. acknowledges Feng Long for valuable feedback and suggestions to the manuscript and Giovanni Rosotti, Stefano Facchini, Giuseppe Lodato and Claudia Toci for fruitful discussion.
E.R. acknowledges financial support from the European Union's Horizon Europe research and innovation programme under the Marie Sk\l{}odowska-Curie grant agreement No. 101102964 (ORBIT-D).
E.R., E. L. and G.L. acknowledge financial support from the European Research Council (ERC) under the European Union’s Horizon 2020 research and innovation programme (grant agreement No. 864965, PODCAST).
This project has received funding from the European Union's Horizon 2020 research and innovation programme under the Marie Sk\l{}odowska-Curie grant agreement No 823823 (DUSTBUSTERS).
CL acknowledges financial support from the UK Science and Technology research Council (STFC) via the consolidated grant ST/W000997/1.
The simulation performed for this paper used the DiRAC Data Intensive service at Leicester, operated by the University of Leicester IT Services, which forms part of the STFC DiRAC HPC Facility (www.dirac.ac.uk).
Fig. \ref{fig:sim} was created using \textsc{splash} \citep{price07a}. All the other figures were created using \textsc{matplotlib} python library \citep{hunter2007}.
\end{acknowledgements}

%
%
\bibliographystyle{aa}
\bibliography{biblio}

\appendix

\section{Other useful pieces of information about eccentric discs}\label{appendix:moreinfo}

\subsection{Eccentric disc eigenmodes}\label{appendix:eccdiscdyn}

In this section, we discuss the main aspects relating to the evolution of the eccentricity $e(a,t)$ and pericentre longitude profiles $\varpi(a,t)$ with time.

Under the assumption that no perturbations to the central, point-like, gravitational potential are present and under the assumption that the material in the disc does not self-interact, neither gravitationally nor through hydrodynamical effects (e.g. pressure/viscosity), $e(a)$ and $\varpi(a)$ do not depend on time.

However, for gaseous astrophysical discs the effects of pressure and the presence of additional non-point-like terms of the gravitational potential -- e.g. disc self-gravity, the presence of a second mass such as a binary star or a planet, oblateness of the central mass or general relativistic corrections --
cause the disc to differentially precess at a rate $\omega(a)$, producing a twisted eccentricity pattern and oscillations of the value of $e(a,t)$ and $\varpi(a,t)$. Pressure, disc self-gravity, and viscosity transport these oscillations as waves throughout the disc. 

For the purpose of reducing the complexity of the equations describing the dynamics of eccentric discs, $e(a,t)$ and $\varpi(a,t)$ can be effectively merged in one complex quantity $E(a,t)$ (e.g. \citealp{ogilvie2001,goodchild&ogilvie2006,ogilvie2008}):
\begin{equation}
    E(a,t)=e(a,t){\rm e}^{\im\varpi(a,t)}.
\end{equation} 
Note the typographical difference between the eccentricity $e$ and the mathematical constant ${\rm e}$.

Both linear ($|E|\ll 1$ and $|a E_a|\ll 1$, i.e. $q\ll 1$ in Eq. \ref{eq:J}) and non-linear theory predict the existence of a special set of solutions of the form (e.g., \citealp{teyssandier2016,ogilvielynch2019}):
\begin{equation}
    E(a,t)=e(a){\rm e}^{\im\omega t}.\label{eq:eigenmodform}
\end{equation}

In absence of dissipation, this expression implies that the disc have a stationary $e(a)$ profile that coherently precesses at the rate $\omega_i$ throughout the whole disc. If dissipation is present the complex phase will also include a twist $\varpi(a)$ of the pericentre orientation, however the whole profile still precesses rigidly conserving the shape of the twist. These configurations are called ``eccentric eigenmodes''.

Eccentric eigenmodes are analogous to normal quantum states. In the linear regime, $E(a,t)$ can be described as a linear superposition of eccentric eigenmodes 
\begin{equation}
E(a,t)=\sum_i c_i E_i(a,t)\quad, 
\end{equation}
where $c_i$ are complex coefficients. Out of the linear regime, non-linear eigenmodes can be always identified even though they do not superimpose in the same way.

The coexistence of multiple eccentric eigenmodes precessing at different rates overall produces some peculiar secular oscillations in the shape of the eccentricity profile and precession rate (e.g. see the evolution of the disc eccentricity in \citealp{thun2017} or \citealp{ragusa2018}).

\subsection{Linear theory and eccentric eigenmodes}\label{appendix:eq}

In this section we provide an explanation of the origin of the functional form of eccentric eigenmodes.

In the approximation of linear perturbations from circular orbits, the unperturbed state has $e=0$ implying $a=R$, making it reasonable to develop the equations in standard cylindrical coordinates.

The evolution of $E(R,t)$ in a 3D locally isothermal disc, due to pressure effects only (i.e. no self-gravity, no companion mass) is governed by \citep{ogilvie2008,ogilvie2014,teyssandier2016}:
\begin{align}
\Sigma R^2 \Omega \pd{E}{t} &= \frac{\im}{R}\pd{}{R}\left( \frac{1}{2}\Sigma \cs^2 R^3 \pd{E}{R} \right) + \frac{\im r}{2}\dd{}{R}\left(\Sigma\cs^2 \right)E\nonumber\\
& \quad - \frac{\im}{2R}\pd{}{R}\left(\Sigma \dd{\cs^2}{r} R^3 E\right) + \frac{3\im}{2R}\Sigma\dd{}{R}\left(\cs^2 R^2\right)E.    \label{eq:3diso}
\end{align}

We note that  eccentric eigenmodes profiles of the form $E_i(R,t)=e_i(R){\rm e}^{\im \omega_i t}$ cause Eq. (\ref{eq:3diso}) to reduce to the form of:
\begin{equation}
\label{eq:generalSL}
  \omega_i \mathcal A(R)e_i(R)=\dd{}{R}\left(\mathcal B(R)\dd{e_i}{R}(R)\right) + \mathcal C(R)e_i(R).
\end{equation}
where $\mathcal A(R)$, $\mathcal B(R)$ and $\mathcal C(R)$ are real functions of the disc parameters.

It can be shown that, after performing some change of variables, Eq. (\ref{eq:generalSL}) can be rewritten in the form of the Shr\"odinger equation:
\begin{equation}
  -\dd{^2\Xi_i(x)}{x^2}+V\Xi_i(x)=\mathcal{E}_i\Xi_i(x),\label{eq:schrod}
\end{equation}
where $x$ and $a$ are related by:
\begin{equation}
     \dd{x}{R} = \left[ \frac{\omega_0 \mathcal A(R)}{\mathcal B(R)}\right]^{1/2}.
\end{equation}
Functions $e_i(R)$ and $\Xi_i(x)$ are related by:
\begin{equation}
    \Xi_i(x)=e_i(R)[\mathcal A(R)\mathcal B(R)]^{1/4}.
\end{equation}
The effective energy eigenvalue $\mathcal{E}_i$ is given by:
\begin{equation}
  \mathcal{E}_i=-\frac{\omega_i}{\omega_0},
\end{equation}
where $\omega_0$ relates with the disc thickness $(H/R)_0$ and mean motion $\Omega_0$ at the inner disc edge as follows:
\begin{equation}
    \omega_0=\frac{1}{2}\left( \frac{H}{R}\right)_0\Omega_0.
\end{equation}
The effective potential $V(x)$ reads:
\begin{equation}
\label{eq:vschrod}
  V=-\frac{\mathcal C(R)}{\omega_0\mathcal A(R)}+(\mathcal A(R)\mathcal B(R))^{-1/4}\dd{^2}{x^2}[\mathcal A(R)\mathcal B(R)]^{1/4}.
\end{equation}

As a consequence, in analogy to quantum mechanics, there exist a countable infinity of eccentric eigenmodes $E_i(R,t)$. Prograde ($\omega_i$) modes correspond to the finite number of negative energy bound states of the potential. These solutions have the form of Eq. (\ref{eq:eigenmodform}), i.e. they feature a stationary eccentricity profile precessing at a rate $\omega_i$.

\subsection{Functional form of the eccentricity profile in circumbinary discs}

Following on from the previous section, we discuss the radial eccentricity profile of eccentric eigenmodes. We solve Eq. (\ref{eq:3diso}) assuming radial power-law profiles of surface density $\Sigma\propto R^{-\sigma}$ and soundspeed $\cs^2 \propto R^{-q}$. Under this assumption, it can be shown that $E(R,t)$ can be expanded in series as \citep{goodchild&ogilvie2006,ogilvie2008}
\begin{equation}
 E(R,t) = \sum_{i} a_i(t) \Psi_i (R) ,
\end{equation}
where $\{a_i(t)\}$ are complex coefficients, which depend on time only. The functions $\Psi_i(R)$ are functions of radius only and read 
\begin{equation}
 \Psi_i(R) = R^{\sigma/2 + q - 1} Z_{\nu} \left( (R/R_i)^{\beta} \right),
\end{equation}
where $Z_{\nu}$ is a form of Bessel function that can be written as
\begin{equation}
 Z_{\nu} (x) = A J_{\nu} (x) + B Y_{\nu} (x) ,
\end{equation}
where  $J_{\nu}$ and $Y_{\nu}$ are Bessel functions of the first and second kind, respectively, with order $\nu$ which is related to $q$, $\sigma$ and $\beta$ by $\nu^2 = \beta^{-2}(2 q + q \sigma + q^2 + \sigma^2/4 - 5)$. $A$, $B$ and $\{ R_i\}$ are set by the (zero gradient) boundary conditions, $\partial_{R} \Psi_i |_{a_{\rm in}} = \partial_{R} \Psi_i |_{a_{\rm out}} = 0$ and an appropriate normalisation condition. 

In a locally isothermal, 3D, powerlaw, disc with only pressure forces then $\beta = (1 + 2 q)/4$ and $\Psi_i$ correspond to the linear eccentric eigenmodes (with $\Psi_i=e_i(R)$ as defined in Eq. \ref{eq:eigenmodform}).  The fundemental mode is then given by
\begin{equation}
 E_0(R,t) =  a_{0}(t) \Psi_{0} (R) .
\end{equation}

When a quadrupole potential (e.g. from a binary), and nonlinear effects are included the fundermental mode will contain contributions from the other $\Psi_i$'s. 
However, at order zero, the radial part of the fundamental mode can be written in the form
\begin{equation}
 e(R) =\mathcal A R^{-\gamma}\left\lbrace  J_{\nu} \left(\left[R/R_0\right]^{\beta}\right) + \mathcal C Y_{\nu} \left(\left[R/R_0\right]^{\beta}\right)\right\rbrace,\label{eq:fundamentalmode}
\end{equation}
where $\mathcal C$, $\beta$, $\gamma$ and $R_0$ are constants whose value depend on the powers of the profiles of $\Sigma$ and $\cs$, as discussed above, as well as other disc properties and other assumptions (e.g. boundary conditions), while $\mathcal A$ sets the amplitude of the mode. 

A simpler functional form can be found by solving Eq. (\ref{eq:3diso}) by using the WKB approximation \citep{shi2012,lee2019b,munoz2020}. Again, assuming radial power-law profiles of surface density and soundspeed, it can be shown that the eccentricity profile in a disc has the form
\begin{equation}
    e(R)=\mathcal A R^{-g}\exp\left[\left(R/r_0\right)^{b}\right].
\end{equation}
Similarly to Eq. (\ref{eq:fundamentalmode}), $b$, $g$ and $r_0$ are constants whose value depend on the powers of the profiles of $\Sigma$ and $\cs$ as well as other disc properties, while $\mathcal A$ sets the amplitude of the mode.

\section{Vertical Structure including bulk viscosity}\label{appendix:bulk}

It is possible to include bulk viscosity for calculating the disc vertical structure, as follows: we introduce the $T^{zz}$ bulk viscous stress tensor term Eq. (\ref{eq:vertvel1}), which becomes:
\begin{equation}
\pd{v^z}{t}+v^a\pd{v^z}{a}+v^\phi\pd{v^z}{\phi}+v^z\pd{v^z}{z}=-\frac{GM}{R^2}\frac{z}{R}-\frac{1}{\rho}\pd{}{z}\left(p-T^{zz}\right).\label{eq:vertvel1bulk}
\end{equation}
where $T^{zz}$ is the bulk viscous stress tensor
\begin{equation}
    T^{zz}=\mu_b\left(\frac{1}{J}\pd{}{\phi}(J\Omega)+\pd{v^z}{z}\right),
\end{equation}
parametrised by a $\mu_b$ given by \citep{ogilvie2001,ogilvie2014,lynch2021}:
\begin{equation}
    \mu_b=\alpha_bp\Omega_0^{-1}.
\end{equation}
The addition of the $T^{zz}$ term causes Eq. (\ref{eq:veloverteq32}) to become
\begin{align}
\Omega^2\frac{\partial ^2 H}{\partial \phi^2}&=-\Omega\pd{H}{\phi}\pd{\Omega}{\phi}-\frac{GM}{R^3}H+\nonumber\\
&\quad\quad\quad +\frac{p_0}{\rho_0 H}\left[1-\frac{\alpha_b}{\Omega_0}\left(\frac{1}{J}\pd{}{\phi}(J\Omega)+\frac{\Omega}{H}\pd{H}{\phi}\right)\right].\label{eq:veloverteq32bulk} 
\end{align}

We note that in order to produce the same plots for $\alpha_b\neq 0$ one needs specify $q$ and $\alpha$ in Eq. (\ref{eq:qcosa}) and (\ref{eq:qsina}) making the parameter space larger. Since the solution would not be general, we do not provide an example plot. 

In practice, the main effect of the bulk viscosity term is an azimuthal shift of the location of the minimum and maximum of $H$ with respect to the pericentre and apocentre of the orbit, where minimum and maximum of $H$ are located for $\alpha_b=0$, respectively. In addition, bulk viscosity produces also small changes in the maximum and minimum vertical velocity reached along the orbit. 

\section{A more general choice of the form of $\rho(a,\phi,z)$ and $p(a,\phi,z)$}\label{appendix:separable}

It can be shown that the final form of Eq. (\ref{eq:veloverteq31}) and (\ref{eq:veloverteq32})  does not change for different assumptions on $\rho(a,\phi,z)$ and $p(a,\phi,z)$, with the only caveat that the form of $\rho$ and $p$ must guarantee the separability of the solution in the variables in the midplane $(a,\phi)$, or $(R,\phi)$, and $z$ (see e.g. \citealp{ogilvie2014}). 

\begin{align}
\rho(a,\phi,z)&=\rho_0(a,\phi)F_\rho(z/H),\\
p(a,\phi,z)&=p_0(a,\phi)F_p(z/H),\\
\pd{}{z}F_p(z/H)&=-\frac{z}{H^2}F_\rho(z/H),
\end{align}
which are fully satisfied when the disc is assumed to be locally isothermal as we did in Eq. (\ref{eq:rho}) and (\ref{eq:p}).

\section{Channel maps and moment 1 maps}
\label{sec:channelmaps}

\begin{figure*}
   \centering
   \includegraphics[width=\textwidth]{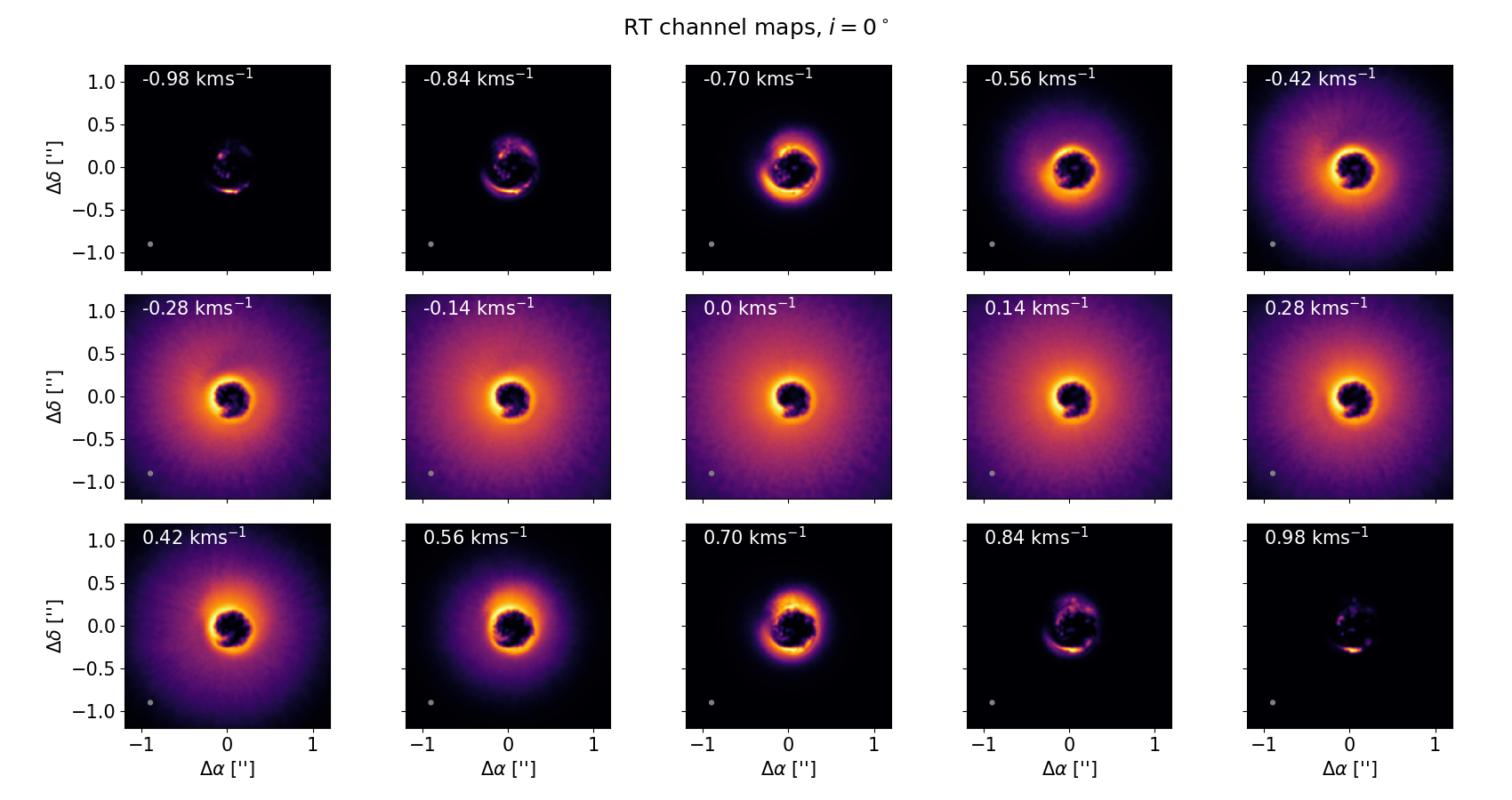}\\
   \includegraphics[width=\textwidth]{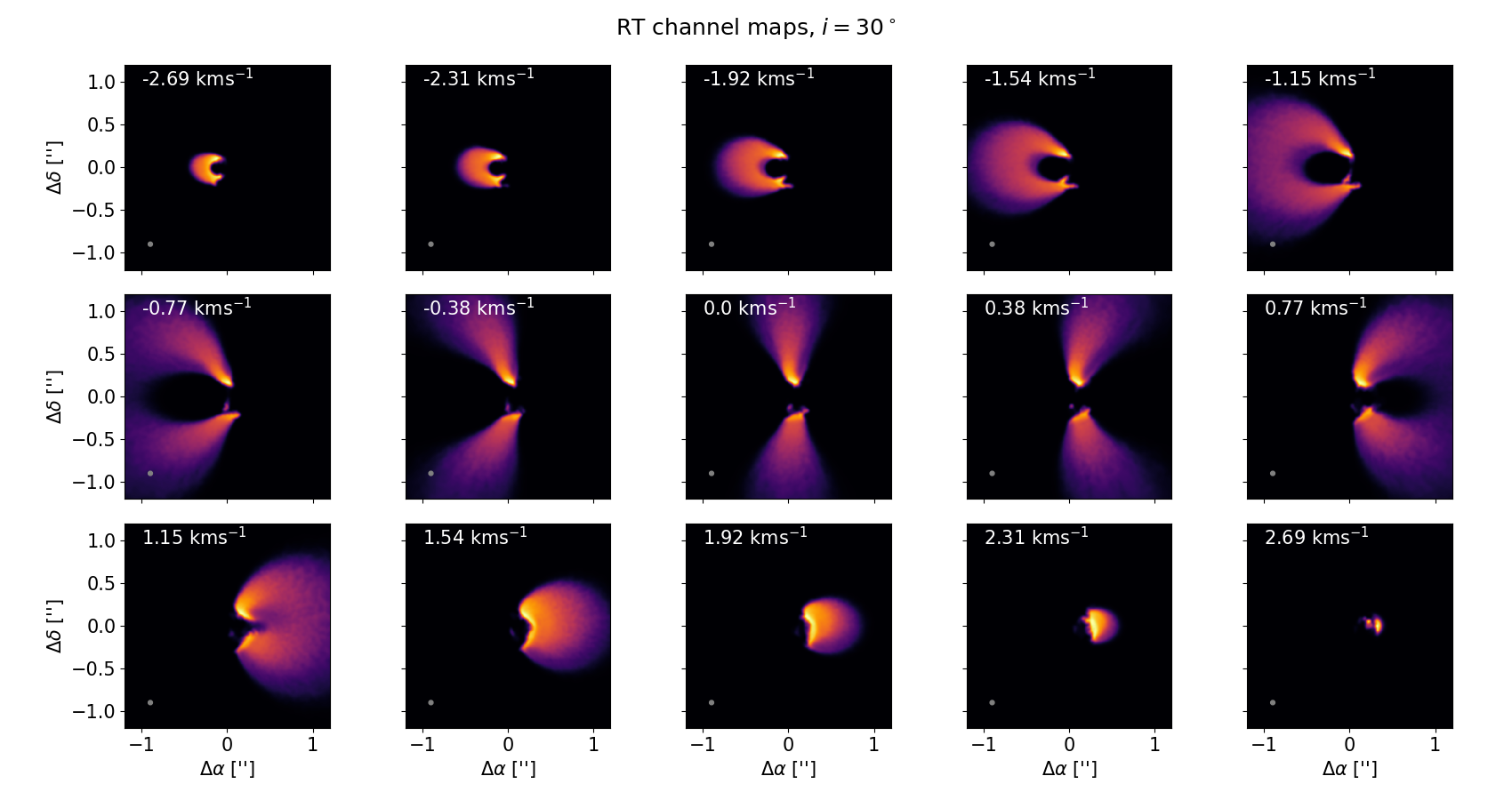}
   \caption{Selection of channel maps from the RT model for the inclination $i=0^\circ$ (top panels) and $i=30^\circ$ (bottom panels), after smoothing the synthetic image from \textsc{mcfost} using a circular Gaussian beam with $\Delta x=0.05''$ and $\Delta v=0.05 \,{\rm kms}^{-1}$. The velocity of each channel is marked in the top left corner of each panel. The grey circle in the bottom left corner represents the corresponding beam size.}
\label{fig:channelmaps}%
\end{figure*}

For completeness, we present in Fig. \ref{fig:channelmaps} a selection of channel maps obtained from the RT simulation for both the $i=0^\circ$ and $i=30^\circ$. Both channel maps are obtained smoothing the ``infinite'' resolution synthetic image from \textsc{mcfost} using a circular Gaussian beam with $\Delta x=0.05''$ and $\Delta v=0.05 \,{\rm kms}^{-1}$.

\end{document}